\documentclass{article}

\usepackage{xcolor}
\newif\ifreview
\ifreview
  \newcommand{\hlyellow}[1]{{\color{red!70!black}#1}}
\else
  \newcommand{\hlyellow}[1]{#1}
\fi

\usepackage{PRIMEarxiv}
\usepackage[utf8]{inputenc}
\usepackage[T1]{fontenc}
\usepackage{hyperref}
\usepackage{url}
\usepackage{booktabs}
\usepackage{nicefrac}
\usepackage{microtype}
\usepackage{fancyhdr}
\usepackage{float}
\usepackage[font=small,labelfont=bf]{caption}
\usepackage{enumitem}

\usepackage{amsmath, amsfonts, amssymb, amsthm, mathtools}
\usepackage{dsfont}
\usepackage{bm}
\newcommand{\degree}{\ensuremath{^\circ}}

\usepackage{graphicx}
\usepackage{tikz}
\usepackage{tikz-cd}
\usepackage{pgfplots}
\pgfplotsset{compat=1.18}

\usepackage[nameinlink,noabbrev]{cleveref}

\theoremstyle{definition}

\theoremstyle{plain}

\theoremstyle{remark}

\newcommand{\cat}[1]{\mathsf{#1}}
\newcommand{\Dyn}{\cat{Dyn}}
\newcommand{\Nat}{\cat{Nat}}
\newcommand{\Art}{\cat{Art}}
\newcommand{\Spec}{\cat{Spec}}
\newcommand{\Comp}{\cat{Comp}}
\newcommand{\id}{\mathrm{id}}

\title{A Category-Theoretic Framework from Biological Mechanics to Engineered Stimulus-Response Systems}

\author{
  \textbf{Lee Marom}$^{1,2}$ \quad
  \textbf{Skylar Tibbits}$^{2}$ \quad
  \textbf{Gioele Zardini}$^{3,4}$ \quad
  \textbf{Markus J. Buehler}$^{1,4,5,*}$ \\[0.6em]
  $^{1}$Department of Mechanical Engineering,\\
  $^{2}$Department of Architecture,\\
  $^{3}$Laboratory for Information and Decision Systems,\\
  $^{4}$Department of Civil and Environmental Engineering,\\
  $^{5}$Center for Computational Science and Engineering, Schwarzman College of Computing, \\
  Massachusetts Institute of Technology, Cambridge, MA, USA \\[0.4em]
  Corresponding author: $^{*}$\texttt{mbuehler@mit.edu}, https://orcid.org/0000-0002-4173-9659
}

\date{}

\begin{document}
\maketitle

\begin{abstract}
Natural materials achieve adaptive behavior through hierarchical organization and coupled mechanisms across scales. Their translation into engineering, however, remains largely heuristic. What is missing is a formal translation framework that carries biological design logic into engineered realization while preserving physical consistency across levels of abstraction. Here we present a category theoretic compositional framework for verified nature-derived design. The framework defines a category of stimulus response dynamical systems with natural and artificial subcategories. It introduces a structure preserving implementation functor from biological mechanics to engineered systems. It also formalizes a machine agnostic specification layer that links behavioral intent to executable fabrication programs. We instantiate the framework on the hygromorphic pinecone hierarchy as a representative biological case. We implement the full pipeline in Grasshopper, where formal specifications are translated into modular parametric scripts that preserve the compositional structure of the model. The resulting designs are fabricated by fused filament fabrication, evaluated experimentally, and tested against model predictions derived from the pipeline. The current implementation generates four actuator classes spanning two stimulus types and two kinematic responses. One actuator arises purely through composition from previously validated components, without additional manual derivation. The results show that compositionality can function not just as a descriptive language, but as a generative and system level verifiable method for mechanical material design. More broadly, the work provides a concrete route for embedding formal multiscale reasoning within increasingly computational, generative, and physics-driven design workflows.
\end{abstract}


\section{Introduction}
\label{sec:intro}
Natural systems exhibit adaptive behaviors that remain difficult to reproduce systematically in engineered artifacts. 
Wheat awns propel seeds into the ground through humidity cycles~\cite{Elbaum2007}, nacre stiffens under impact~\cite{Cartwright2007}, and a Venus flytrap snaps shut under mechanical contact~\cite{Forterre2005}. 
In each case, the macroscopic response emerges from interacting stimulus-response mechanisms organized across multiple length scales~\cite{Wegst2015, Barthelat2016}. 
Translating such behavior into an engineered system requires preserving not only the mechanisms operating at each scale, but also the interfaces that connect them.

This challenge has motivated a large body of work in bioinspired and biomimetic design~\cite{Mohammed2009}. 
Yet most translations still proceed by case-based analogy: a designer identifies a qualitative parallel between a biological mechanism and an engineering material or geometry, then builds an artifact intended to mimic the observed form or response. 
This strategy has produced striking demonstrations, but it becomes increasingly fragile as the target behavior becomes multiscale and compositional. 
When function arises not from a single mechanism, but from interactions among mechanisms across scales, analogy alone offers no guarantee that the assembled artifact will reproduce the assembled behavior. 
Each scale interface must be revalidated independently, and the space of possible translations grows combinatorially with hierarchy depth.

The recognition that design problems have compositional structure has a long history. 
Early work in design theory framed the designer's task as achieving fit between form and context, arguing that as problems grow in complexity they must be represented through the interactions among their constituent variables and decomposed into hierarchically nested subsets of weakly interacting elements~\cite{Alexander1964}. 
What has been missing is a mathematical language that makes this compositional structure formal and verifiable.

Category theory has emerged as a mathematical language for compositional structure in engineering and science~\cite{Spivak2014, censi2022}. 
Its central premise is that complex systems are best understood not through their internal states but through how their parts compose. 
A category consists of objects and structure-preserving maps (called morphisms) between them, together with a composition rule that guarantees associativity and the existence of identities. 
A functor is a map between two categories that preserves this compositional structure, carrying objects to objects and morphisms to morphisms in a way that respects how they connect.
These constructs have the potential to formalize what biomimetic analogy cannot guarantee. 
If each interface in a hierarchy is verified locally, the assembled system is verified globally, because composition of verified maps is itself verified. 
Two systems that share the same compositional structure, such as a biological material hierarchy and an engineered one, can be related by a functor that preserves the interface logic at every scale without requiring that the two domains share any physical substrate.

Applied category theory has been developed across several domains. 
In knowledge representation, \textit{ologs} use functors to translate meaning between conceptual frameworks while preserving the relations among concepts~\cite{Spivak2012}. 
In dynamical systems, compositional semantics for interconnected open systems shows that the behavior of an assembly can be derived from the behaviors of its parts together with the way they are wired~\cite{Schultz2020}. 
In engineering design, co-design theory formalizes the simultaneous design of interconnected subsystems and shows that feasibility of a global specification can be computed compositionally from the feasibility of its components~\cite{Censi2015, Zardini2021, zardini2023co}. 
In physics, categorical quantum mechanics reformulated quantum theory in terms of process composition and recovered substantive physical content from compositional structure alone, showing that focusing on structure and relations can clarify a theory whose standard formalism obscures its compositional logic~\cite{abramsky2009}. 
These works demonstrate that categorical and compositional methods can bring formal clarity to problems whose complexity arises from the interaction of parts.

Within materials science, category theory has been used to formalize the building-block replacement problem in hierarchical material design, determining which substitutions of components at one scale preserve function at the next~\cite{Giesa2012}. 
Separately, functorial mappings between categories revealed structural similarities across domains as distant as hierarchical protein materials and musical compositions~\cite{Giesa2011}. Categorical prototyping extended these ideas to additive manufacturing by creating fabrication protocols inspired by molecular-scale mechanisms~\cite{Brommer2016}. 
The same emphasis on objects and relations appears in recent AI-driven materials design, where ontological and graph-based frameworks guide generative search over material architectures~\cite{Buehler2024, Buehler2025}. This relational view also extends into neural architectures. Multi headed self attention has been interpreted as a graph forming realization of \textit{ologs}, enabling prediction of multiscale physical fields and material properties
without convolutional priors~\cite{Buehler2022FieldPerceiver}. 
Coupling these generative efforts with a formal understanding of how biological materials encode stimulus-response logic across scales~\cite{Marom2025} is a necessary step toward systematic translation of nature-derived design into engineered systems.

Three features distinguish this construction from prior categorical treatments of hierarchical materials. First, the framework is closed end-to-end. It connects biological mechanics to machine-executable fabrication through a single chain of structure-preserving maps,
whereas earlier categorical work addressed substitution, analogy, or
prototyping as isolated facets. Second, the translation is
dynamical rather than purely structural. Objects of $\mathsf{Dyn}$ carry
evolution laws and morphisms satisfy a simulation condition. What
is preserved across scales is therefore the stimulus-response behavior itself,
not just the connectivity of building blocks. Third, the framework
is verified by construction. Closure of the simulation condition
under composition guarantees that any chain of locally valid scale
transitions is itself valid. The projection
$\pi : \mathsf{Spec} \to \mathsf{Art}$ makes this guarantee
operational by separating behavioral content from the fabrication
annotations that realize it. \hlyellow{Here, verification is a claim about composition, not about any single model. Given valid local models and interfaces, the framework guarantees that their composition stays consistent across scales. The fidelity of the models themselves, together with material variability and fabrication tolerances, remains empirical and is assessed through physical validation.} Together these features shift the
categorical translation of nature-derived design from a descriptive
stance to a generative and verifiable one.

\hlyellow{The mechanics used at each scale enter as exchangeable components, and any local model can be substituted as long as its interfaces compose. The contribution is therefore a general method for composition and verification, not a commitment to the specific models realized here.}

The full pipeline chains four categorical layers, each connected by a structure-preserving map:
\begin{equation}
\underbrace{
  F_{\mathrm{fib}}
  \xrightarrow{\;\alpha_1\;}
  L_{\mathrm{lam}}
  \xrightarrow{\;\alpha_2\;}
  T_{\mathrm{tis}}
  \xrightarrow{\;\alpha_{\mathrm{geom}}\;}
  S_{\mathrm{elem}}
  \xrightarrow{\;\alpha_3\;}
  O_{\mathrm{org}}
}_{\displaystyle\Nat}
\;\xrightarrow{\;\mathcal{F}\;}
\underbrace{A}_{\displaystyle\Art}
\;\xleftarrow{\;\pi\;}
\underbrace{\Sigma}_{\displaystyle\Spec}
\;\xrightarrow{\;\mathcal{E}\;}
\underbrace{C}_{\displaystyle\mathbf{Comp}}.
\end{equation}
The following sections develop this formalization as a general framework for nature-derived material design. We use the pinecone hygromorphic hierarchical stimulus-response system as a running example on how to implement the compositional framework for a given natural system in $\Nat$. We use 4D printed bilayer composites as the engineered analog in $\Art$. Finally, we validate the resulting designs through fabrication and testing.

The framework we develop can be read as a formal reconstruction of what bioinspired design or biomimicry has always aspired to, translating the organizing logic of nature into engineered form, but in a setting where every interface in that translation is made mathematically explicit and computable, and the resulting artifact is verified by the structure of the translation itself.

\section{Formalizing stimulus-response systems}
\label{sec:dyn}
Adaptive behavior in biological and engineered materials arises from stimulus-response mechanisms operating across multiple length scales. 
A cellulose fiber swells under humidity. 
A printed polymer filament bends under moisture uptake. A shape-memory wire contracts under heat. 
Despite their different physics, these systems share the same abstract structure: an internal state, an environmental stimulus, and a governing law that determines how the state evolves.
This structure recurs throughout a material hierarchy, from molecular constituents to macroscopic organs.
If the overall behavior is to remain consistent, the maps connecting one scale to the next must preserve it.

We formalize this recurring structure through the category $\Dyn$ of stimulus-response dynamical systems. 
A category, in the sense used throughout this paper, is a collection of objects together with structure-preserving maps (morphisms) between them that compose associatively and admit identities.
In $\Dyn$, each object represents a system at a given scale, each morphism represents a map between systems that preserves their dynamics, and composition encodes the chaining of scale transitions across a hierarchy. 
The subcategories $\Nat$ and $\Art$ distinguish natural from engineered systems within this shared formal setting.

An object of $\Dyn$ is a triple
\begin{equation}
\label{eq:dyn-obj}
S := (X,\, E,\, f),
\end{equation}
where $X$ is a finite-dimensional state space, $E$ is a finite-dimensional environment (stimulus) space, and $f$ is the governing law that determines how the state evolves under a given stimulus. The system evolves according to dynamics
\begin{equation}
\label{eq:dyn-ode}
\dot{x} = f(x,e).
\end{equation}

A morphism $(\alpha, \alpha_E) : (X,E,f) \to (Y,F,g)$ in $\Dyn$ is a pair of smooth maps, one on states ($\alpha : X \to Y$) and one on stimuli ($\alpha_E : E \to F$), satisfying the simulation condition
\begin{equation}
\label{eq:dyn-diff}
d\alpha_x\big(f(x,e)\big) = g\big(\alpha(x),\, \alpha_E(e)\big)
\end{equation}
for all $(x,e) \in X \times E$, where $d\alpha_x$ is the differential of $\alpha$ at $x$. 
In words, the map commutes with time evolution: evolving at the fine scale and then mapping through $\alpha$ gives the same result as first mapping and then evolving at the coarse scale.

Throughout this paper, morphisms in $\Dyn$ play two recurring roles. Assembly morphisms, denoted $\alpha$, aggregate fine-scale subsystems into coarser descriptions. 
Reduction morphisms, denoted $\beta$, extract the observables needed at the next scale or at the final measured output. 
\hlyellow{In this usage, $\beta$ may denote either a dynamical reduction, when the extracted variables are passed forward with an induced evolution law, or a terminal observable, when the extracted quantity is used only as a final measured or geometric output (Appendix~\ref{appendix}).}
Morphisms compose component-wise. 
For composable pairs $(\alpha, \alpha_E)$ and $(\beta, \beta_F)$, the composite $(\beta \circ \alpha,\, \beta_F \circ \alpha_E)$ again satisfies the simulation condition (equation~\ref{eq:dyn-diff}). 
This closure under composition is what allows scale transitions to be chained consistently across a hierarchy. 
Formal definitions for categories and the proof that $\Dyn$ is a category are given in Appendix~\ref{appendix}.

$\Nat$ is the subcategory of $\Dyn$ whose objects arise from natural or biological mechanisms. 
$\Art$ is the subcategory whose objects are realized as engineered or fabricated systems. 
Both inherit the categorical structure of $\Dyn$.
The central goal of this paper is to construct a scale-by-scale translation from a hierarchy in $\Nat$ to a structure-preserving counterpart in $\Art$, where the engineered system at each scale shares the same state-space organization and evolution law as its biological counterpart, with differences entering through the parameters of the engineered realization. 
This translation is formalized as a functor $\mathcal{F}:\Nat\to\Art$ and is developed in Section~\ref{sec:nat-art} after the biological hierarchy is instantiated.

\section{Biological hierarchy and governing mechanics}
\label{sec:nat}

The subcategory $\Nat \subset \Dyn$ contains stimulus-response systems arising from biological mechanisms. 
It inherits the objects, morphisms, and composition structure of $\Dyn$ (Appendix~\ref{appendix}) and provides the formal setting in which a biological material hierarchy can be written as a chain of coupled dynamical systems.
The Venus flytrap achieves rapid closure through a turgor-driven change in leaf curvature that triggers a bistable elastic snap~\cite{Forterre2005}. Wheat awns exploit cellulose fibril arrangements that convert humidity cycles into directed bending, propelling seeds into the ground~\cite{Elbaum2007}. Nacre assembles a hierarchical composite from molecular to macro scale through controlled self-assembly of mineral platelets in an organic matrix~\cite{Cartwright2007}. Each of these can be represented as a chain of objects in $\Nat$.

\begin{figure}[ht]
    \centering
    \includegraphics[width=0.85\linewidth]{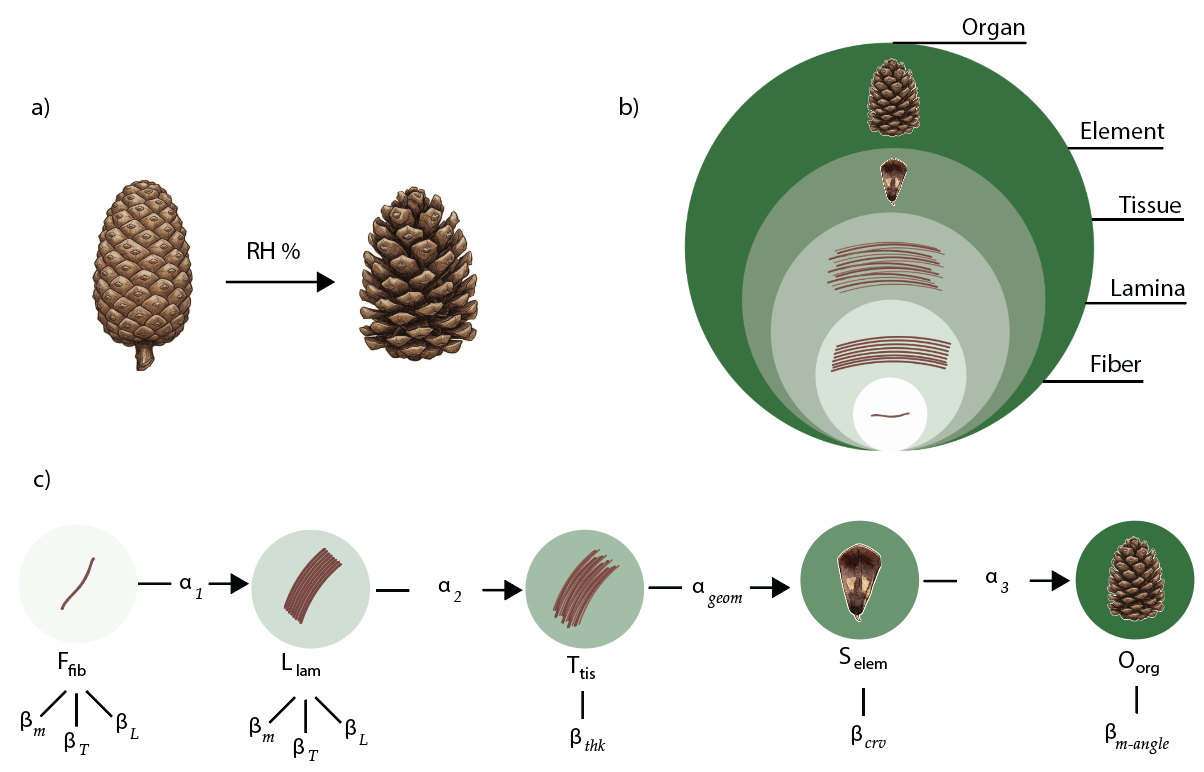}
    \caption{The pinecone hierarchy in $\Nat$. (a) Dynamic pinecone opens and closes in response to relative humidity (RH). (b) Compositional multiscale structure  across modeled scales: fiber, lamina, tissue, element, organ. (c) The complete formalization as a chain of objects with assembly morphisms $\alpha$ connecting successive scales and reduction morphisms $\beta$ extracting observables.}
    \label{fig:nat}
\end{figure}

Here we instantiate $\Nat$ on the hygromorphic pinecone hierarchy as an example of a natural system whose governing mechanics are well characterized from cell wall to organ scale~\cite{Zhang2021, Quan2021, Reyssat2009}.
The pinecone hierarchy comprises five objects in $\Nat$, each representing the stimulus-response system at one scale and linked to the next by assembly morphisms that aggregate lower-level descriptions into higher-level ones (Figure~\ref{fig:nat}).
At each scale, reductions $\beta$ extract the observables required at the next interface. The following subsections specify, for each scale, the state space, stimulus, governing law, assembly morphism, and reductions.

\subsection{Fiber scale}
\label{sec:fiber}

The fiber is the smallest subsystem we model at which the stimulus generates a primary strain response. 
In the pinecone, it represents a cell-wall bundle whose moisture uptake produces anisotropic eigenstrain, consistent with established models of wood cell-wall hygromechanics~\cite{Zhang2021}. 
The fiber system is $F_{\mathrm{fib}} := (X_{\mathrm{fib}}, E_{\mathrm{RH}}, f_{\mathrm{fib}}) \in \Nat$.

The fiber state space is
\begin{equation}
  X_{\mathrm{fib}} = \{(m,\varepsilon_L,\varepsilon_T)\} \subset \mathbb{R}^3,
\end{equation}
where $m$ is the moisture content and $(\varepsilon_L, \varepsilon_T)$ are the principal strain components aligned with and transverse to the microfibril direction. The environment is the scalar relative humidity, $E_{\mathrm{RH}} = \{u\} \subset \mathbb{R}$.

Taking the swelling coefficients as independent of humidity~\cite{Reyssat2009}, the eigenstrain is linear in moisture change about a reference value $m_0$. We model the dynamics as a first-order relaxation toward equilibrium,
\begin{equation}
  f_{\mathrm{fib}}(m,\varepsilon_L,\varepsilon_T;u)
  =
  \begin{pmatrix}
    -\tau_m^{-1}(m - m_{\mathrm{eq}}(u)) \\
    -\tau_L^{-1}(\varepsilon_L - \chi_L(m-m_0)) \\
    -\tau_T^{-1}(\varepsilon_T - \chi_T(m-m_0))
  \end{pmatrix},
\end{equation}
where $m_{\mathrm{eq}}(u)$ is the equilibrium moisture content, $\chi_L$ and $\chi_T$ are the longitudinal and transverse swelling coefficients, and $\tau_m$, $\tau_L$, $\tau_T$ are the associated relaxation times~\cite{Zhang2021}.

\begin{figure}[ht]
    \centering   \includegraphics[width=0.23\linewidth]{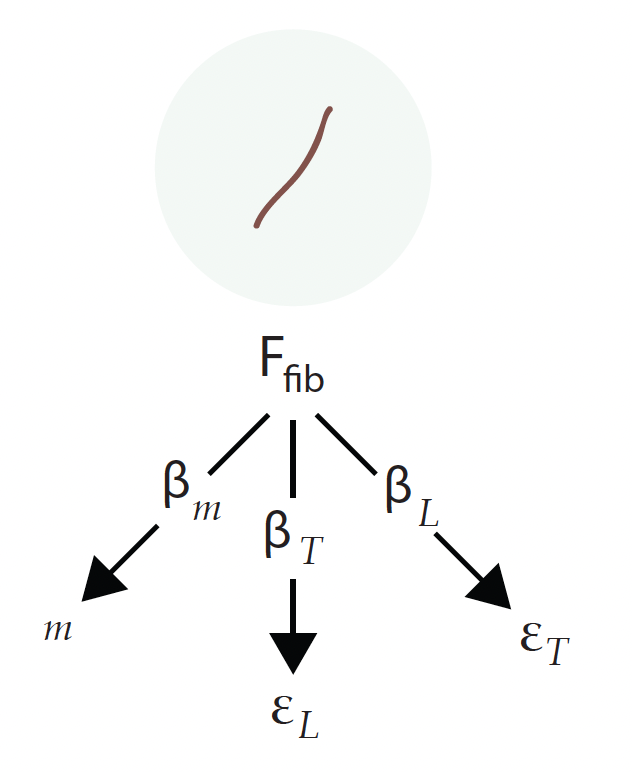}
    \caption{Fiber-scale reductions. The three projections $\beta_m$, $\beta_L$, and $\beta_T$ extract the moisture content $m$ and the longitudinal and transverse strain components $\varepsilon_L$, $\varepsilon_T$ from the fiber state.}
    \label{fig:fib}
\end{figure}

Three reduction morphisms simply project onto the state components (Figure~\ref{fig:fib}),
\begin{equation}
  \beta_m = m, \qquad \beta_L = \varepsilon_L, \qquad \beta_T = \varepsilon_T.
\end{equation}
These provide the moisture and strain observables from which the lamina scale is assembled.

\subsection{Lamina scale}

We model a lamina as a thin, mechanically coherent band of $N$ aligned fibers. 
Because the fibers share a common orientation, the lamina inherits the same principal strain directions $(\varepsilon_L,\varepsilon_T)$ as its constituents. 
The lamina system is $L_{\mathrm{lam}} := (X_{\mathrm{lam}}, E_{\mathrm{RH}}, f_{\mathrm{lam}}) \in \Nat$.

\begin{figure}[ht]
    \centering
    \includegraphics[width=0.45\linewidth]{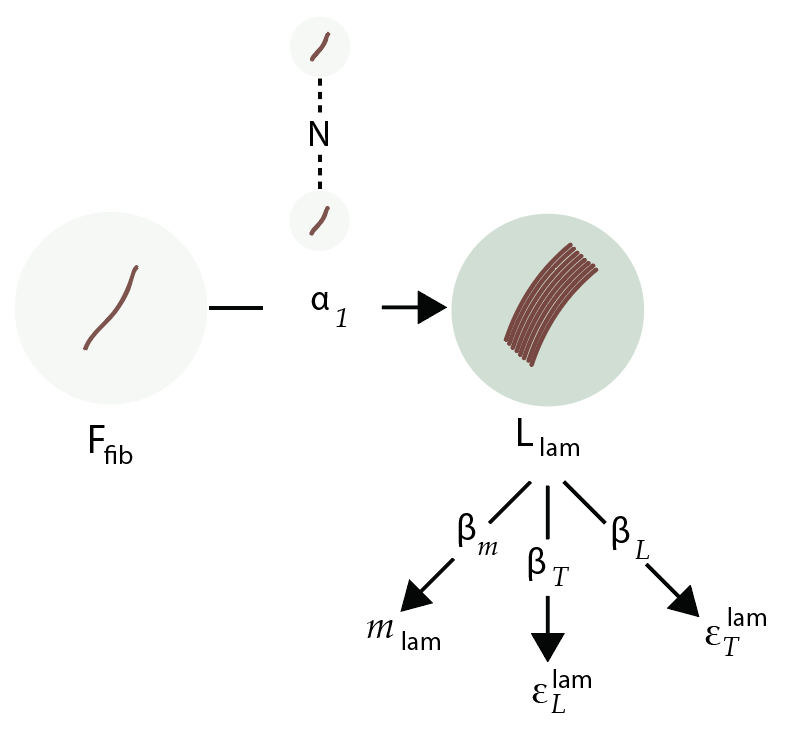}
    \caption{Fiber-to-lamina assembly. The assembly morphism $\alpha_1$ averages $N$ fiber states (dotted arrows) into the lamina state $L_{\mathrm{lam}}$. Reductions $\beta_m^{\mathrm{lam}}$, $\beta_L^{\mathrm{lam}}$, $\beta_T^{\mathrm{lam}}$ extract the lamina-scale observables.}
    \label{fig:lam}
\end{figure}

The assembly morphism $\alpha_1$ averages the $N$ fiber states into effective lamina fields (Figure~\ref{fig:lam}),
\begin{equation}
\alpha_1\big((m_i,\varepsilon_{L,i},\varepsilon_{T,i})_{i=1}^N\big)
=
\left(
\frac{1}{N}\sum_{i=1}^N m_i,\;
\frac{1}{N}\sum_{i=1}^N \varepsilon_{L,i},\;
\frac{1}{N}\sum_{i=1}^N \varepsilon_{T,i}
\right),
\end{equation}
giving the lamina state space $X_{\mathrm{lam}} = \{(m_{\mathrm{lam}}, \varepsilon_L^{\mathrm{lam}}, \varepsilon_T^{\mathrm{lam}})\} \subset \mathbb{R}^3$ with environment $E_{\mathrm{RH}} = \{u\}$. \hlyellow{The map $\alpha_1$ acts on the product of $N$ fiber states, so each of the $N$ inputs takes its own value. In the pinecone modeled here the fibers within a lamina share the same constitutive parameters and the same orientation, consistent with the largely uniform fiber organization within a pinecone scale layer~\cite{Reyssat2009, Quan2021}. Figure~\ref{fig:lam} depicts this uniform case, where the average returns a representative lamina state. The case of fibers carrying differing parameters is identified as a future extension in Appendix~\ref{appendix}.} The lamina evolution law $f_{\mathrm{lam}}$ 
has the same relaxation form as the fiber, evaluated at the averaged state
\begin{equation}
  f_{\mathrm{lam}}(m_{\mathrm{lam}}, \varepsilon_L^{\mathrm{lam}}, 
  \varepsilon_T^{\mathrm{lam}}; u)
  =
  \begin{pmatrix}
    -\tau_m^{-1}(m_{\mathrm{lam}} - m_{\mathrm{eq}}(u)) \\
    -\tau_L^{-1}(\varepsilon_L^{\mathrm{lam}} - \chi_L(m_{\mathrm{lam}}-m_0)) \\
    -\tau_T^{-1}(\varepsilon_T^{\mathrm{lam}} - \chi_T(m_{\mathrm{lam}}-m_0))
  \end{pmatrix}.
\end{equation}

The corresponding reductions, which provide the observables needed to assemble the tissue scale, are
\begin{equation}
  \beta_m^{\mathrm{lam}} = m_{\mathrm{lam}}, \qquad
  \beta_L^{\mathrm{lam}} = \varepsilon_L^{\mathrm{lam}}, \qquad
  \beta_T^{\mathrm{lam}} = \varepsilon_T^{\mathrm{lam}}.
\end{equation}

\subsection{Tissue scale}

The tissue here is described as a multilayer composite obtained by stacking $M$ laminae, each with its own fiber orientation angle $\phi^{(j)}$ relative to a shared tissue coordinate system. In the pinecone, sublayers respond differently to humidity because their microfibril orientations differ~\cite{Zhang2021, Reyssat2009}. It is this difference in orientation between laminae that generates the through-thickness strain mismatch driving macroscopic deformation. The tissue system is $T_{\mathrm{tis}} := (X_{\mathrm{tis}}, E_{\mathrm{RH}}, 
f_{\mathrm{tis}}) \in \Nat$.

\begin{figure}[ht]
    \centering    
    \includegraphics[width=0.6\linewidth]{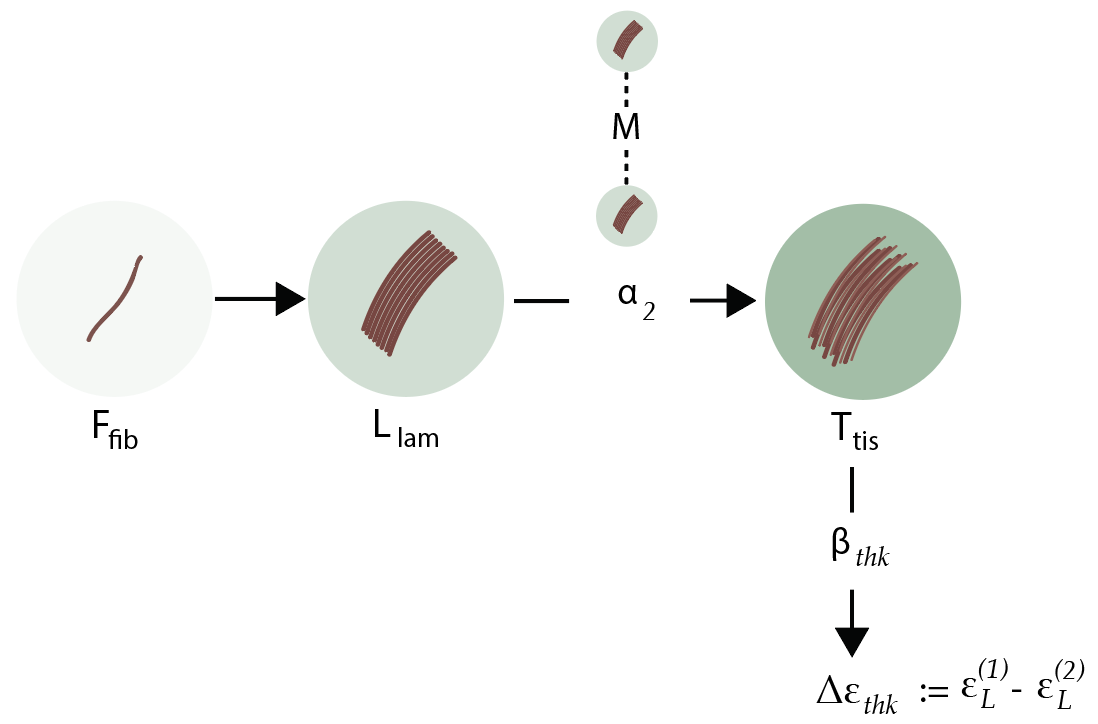}
    \caption{Lamina-to-tissue assembly. The assembly morphism $\alpha_2$ stacks $M$ laminae with different fiber orientations $\phi^{(j)}$ into the tissue state $T_{\mathrm{tis}}$. The reduction $\beta_{\mathrm{thk}}$ extracts the through-thickness strain mismatch $\Delta\varepsilon_{\mathrm{thk}}$.}
    \label{fig:tis}
\end{figure}

The assembly morphism $\alpha_2$ stacks $M$ lamina states into the tissue state space (Figure~\ref{fig:tis}),
\begin{equation}
  X_{\mathrm{tis}} = X_{\mathrm{lam}}^{M}, \qquad
  x_{\mathrm{tis}} = \big(x_{\mathrm{lam}}^{(1)}, \dots, x_{\mathrm{lam}}^{(M)}\big),
\end{equation}
with environment $E_{\mathrm{RH}} = \{u\}$. Each lamina evolves independently under the shared humidity input, giving the tissue evolution law
\begin{equation}
  f_{\mathrm{tis}}(x_{\mathrm{tis}}, u) 
  = \big(f_{\mathrm{lam}}(x_{\mathrm{lam}}^{(1)}, u),\; \dots,\; 
    f_{\mathrm{lam}}(x_{\mathrm{lam}}^{(M)}, u)\big).
\end{equation}

To aggregate the effects of swelling of all laminae within the tissue, we collect the strain of each lamina driven by the fiber orientation angle relative to a defined tissue axis. For a lamina whose fibers are oriented at angle $\phi^{(j)}$ to the tissue axis, the effective longitudinal strain is 
\begin{equation}
\varepsilon^{\mathrm{lam}} = \varepsilon_L^{(j)} \cos^2\phi^{(j)} + \varepsilon_T^{(j)} \sin^2\phi^{(j)}.
\end{equation}

For the bilayer case ($M = 2$) used throughout this paper, the reduction $\beta_{\mathrm{thk}}$ extracts the longitudinal strain mismatch across the thickness:
\begin{equation}
  \beta_{\mathrm{thk}}(x_{\mathrm{tis}}) = \Delta\varepsilon_{\mathrm{thk}} := \varepsilon^{\mathrm{lam}}_1 - \varepsilon^{\mathrm{lam}}_2.
\end{equation}
 When the two laminae have different orientations, they swell by different amounts along the tissue axis and $\Delta\varepsilon_{\mathrm{thk}}$ drives bending curvature. The reduction $\beta_{\mathrm{thk}}$ supplies the curvature-driving eigenstrain to the element scale.

\subsection{Element scale}

The element is the mesoscale unit whose observable deformation emerges from the tissue-level strain mismatch. The element system is $S_{\mathrm{elem}} := (X_{\mathrm{elem}}, E_{\mathrm{RH}}, f_{\mathrm{elem}}) \in \Nat$.

\begin{figure}[ht]
    \centering    \includegraphics[width=0.65\linewidth]{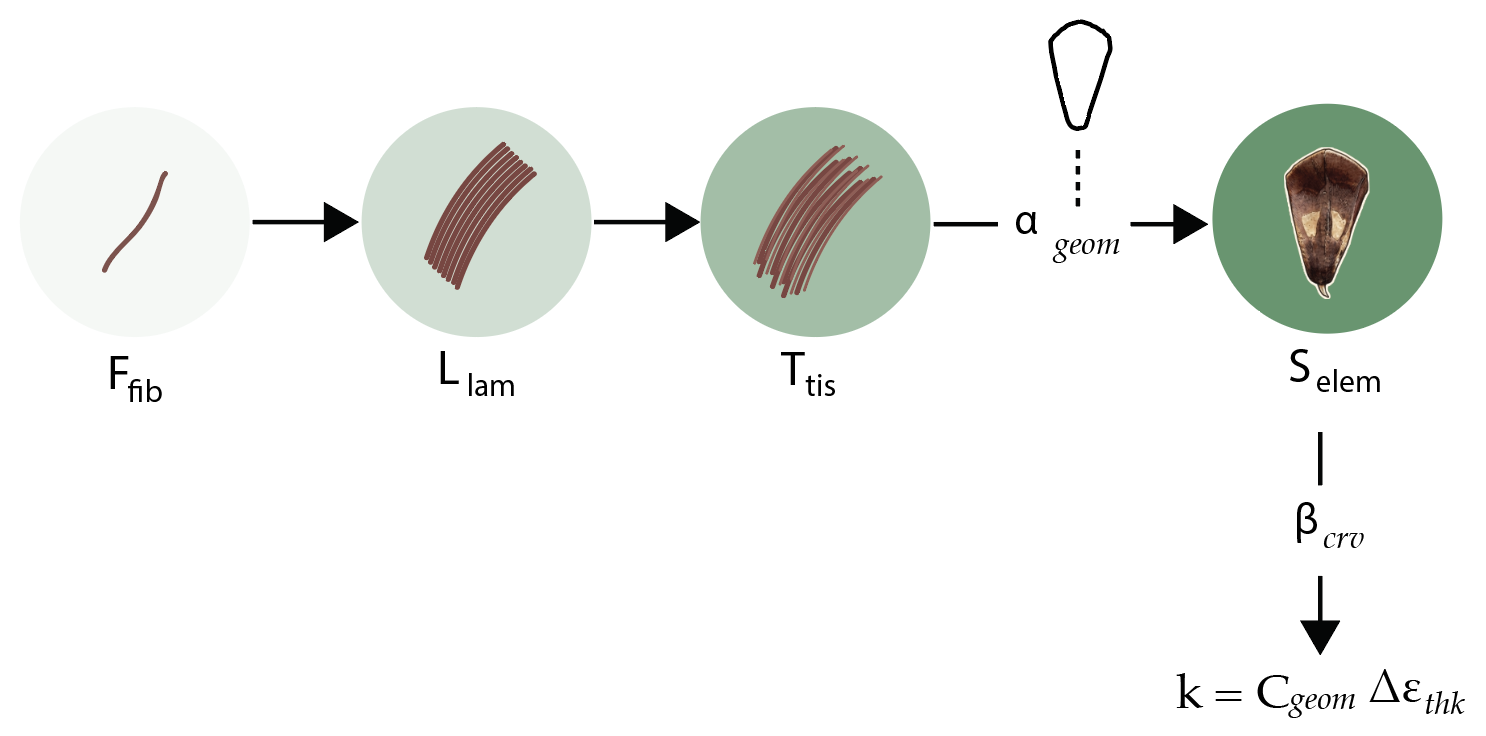}
    \caption{Element scale. The assembly morphism $\alpha_{\mathrm{geom}}$ applies the geometric constraints of the pinecone scale to the tissue, producing the structural element $S_{\mathrm{elem}}$. The reduction $\beta_{\mathrm{crv}}$ extracts the intrinsic curvature $\kappa$.}
    \label{fig:elem}
\end{figure}

The assembly morphism $\alpha_{\mathrm{geom}}$ applies geometric parameters of the structural element (e.g.\ width $w$, length $L$, layer thicknesses $h_1$, $h_2$) to the tissue state, selecting the appropriate kinematic model (Appendix~\ref{appendix}\hlyellow{;} Figure~\ref{fig:elem}). Because pinecone scales satisfy the slender-beam limit ($w/L \ll 1$)~\cite{Reyssat2009}, the Timoshenko bilayer formula~\cite{Timoshenko1925} gives the intrinsic curvature as
\begin{equation}
  \kappa = C_{\mathrm{geom}}\,\Delta\varepsilon_{\mathrm{thk}},
\end{equation}
where the geometric coefficient is
\begin{equation}
\hlyellow{C_{\mathrm{geom}} = \frac{1}{h}\cdot\frac{6(1+\mu)^2}{3(1+\mu)^2 + (1+\mu n)\!\left(\mu^2 + \dfrac{1}{\mu n}\right)}, \qquad \mu = \frac{h_1}{h_2},\quad n = \frac{E_1}{E_2},\quad h = h_1 + h_2.}
\end{equation}

The element state space is $X_{\mathrm{elem}} = \{(\kappa, \Delta\varepsilon_{\mathrm{thk}}, m)\} \subset \mathbb{R}^3$ with environment $E_{\mathrm{RH}} = \{u\}$. Differentiating the curvature relation and substituting the mismatch relaxation gives the element evolution law,
\begin{equation}
  f_{\mathrm{elem}}(\kappa,\Delta\varepsilon_{\mathrm{thk}},m;u)
  =
  \begin{pmatrix}
    -C_{\mathrm{geom}}\,\tau_{\Delta}^{-1}
      (\Delta\varepsilon_{\mathrm{thk}}-\Delta\varepsilon^{0}(m))
    \\[4pt]
    -\tau_{\Delta}^{-1}
      (\Delta\varepsilon_{\mathrm{thk}}-\Delta\varepsilon^{0}(m))
    \\[4pt]
    -\tau_m^{-1}(m-m_{\mathrm{eq}}(u))
  \end{pmatrix},
\end{equation}
where $\tau_{\Delta}$ is the mismatch relaxation time and $\Delta\varepsilon^{0}(m)$ is the equilibrium mismatch strain. 

The reduction $\beta_{\mathrm{crv}}$ extracts the intrinsic curvature $\kappa$ from the element state, supplying the observable from which the organ-scale opening angle is computed.

\subsection{Organ scale}

The organ is described as a structural assembly of $K$ elements evolving under a shared stimulus. Each element carries its own intrinsic curvature $\kappa_i$, and its contribution to the macroscopic shape is determined by its attachment to the organ. In the pinecone, each scale is clamped at its base to a central core, with free length $L_i$ extending beyond the attachment. The organ system is $O_{\mathrm{org}} := (X_{\mathrm{org}}, E_{\mathrm{RH}}, f_{\mathrm{org}}) \in \Nat$.

\begin{figure}[ht]
    \centering    \includegraphics[width=0.8\linewidth]{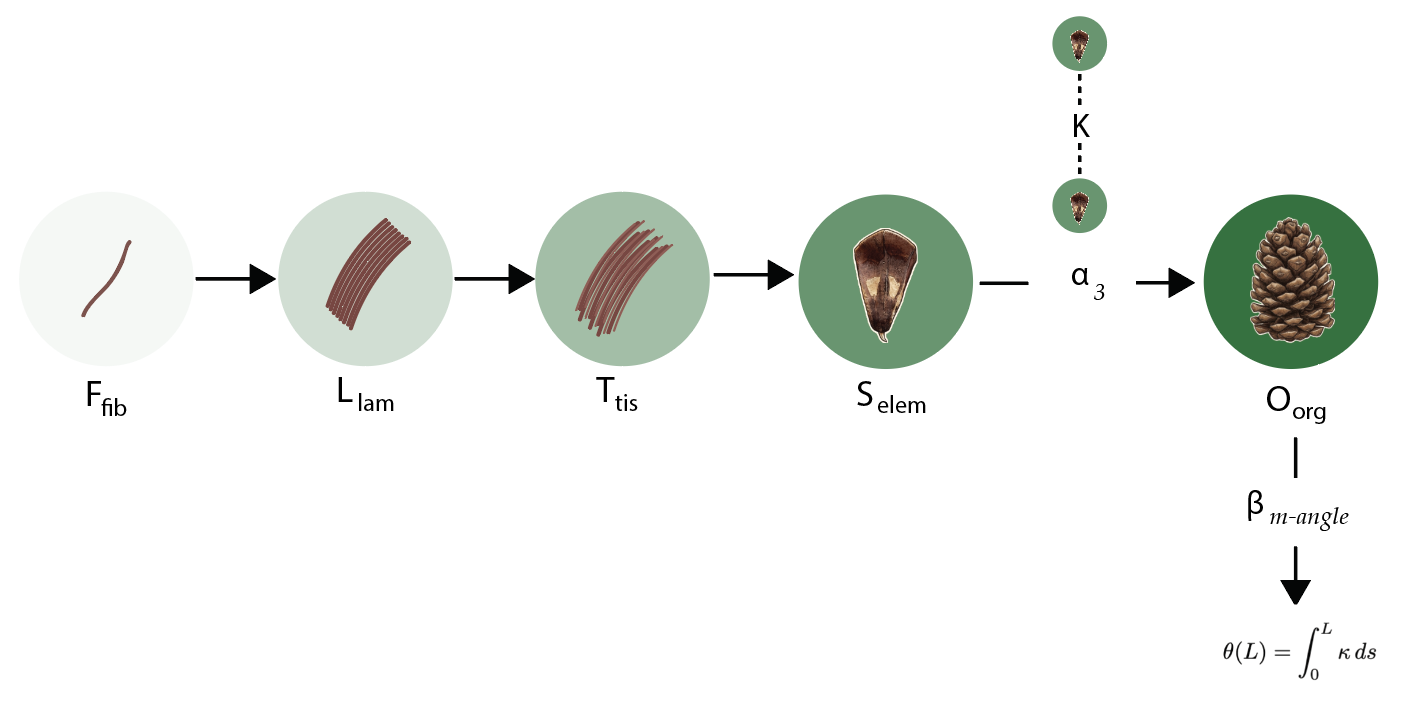}
    \caption{Organ scale. The assembly morphism $\alpha_3$ aggregates $K$ elements (dotted line) into the organ $O_{\mathrm{org}}$. The reduction $\beta_{\mathrm{m\text{-}angle}}$ extracts the mean opening angle $\theta$ as the macroscopic observable.}
    \label{fig:org}
\end{figure}

The assembly morphism $\alpha_3$ embeds each element into a global kinematic frame through these boundary conditions (Figure~\ref{fig:org}). For a scale with uniform curvature $\kappa$ and free length $L$ clamped at $\theta(0) = 0$, the opening angle is
\begin{equation}
  \theta(L) = \int_0^L \kappa \, ds = L\kappa.
\end{equation}
No new constitutive laws are introduced at this scale. Each element evolves  independently under the shared stimulus, giving the organ evolution law
\begin{equation}
  f_{\mathrm{org}}(x_{\mathrm{org}}, u) 
  = \big(f_{\mathrm{elem}}(x_{\mathrm{elem}}^{(1)}, u),\; \dots,\; 
    f_{\mathrm{elem}}(x_{\mathrm{elem}}^{(K)}, u)\big).
\end{equation}

The reduction $\beta_{\mathrm{m\text{-}angle}}$ extracts the mean opening angle across all \hlyellow{$K$} elements as the macroscopic observable.

\section{Compositional translation from biology to engineering}
\label{sec:nat-art}

The subcategory $\Art \subset \Dyn$ contains stimulus-response systems realized as artificial or engineered artifacts. 
Fiber-reinforced bilayers that bend under moisture uptake~\cite{SydneyGladman2016}, shape-memory polymers that recover a target shape under heating~\cite{Lendlein2002}, and self-healing hydrogels~\cite{Taylor2016} are all examples of possible objects in $\Art$. 
Like their biological counterparts in $\Nat$, these systems are characterized by a state space, an environmental stimulus, and a governing evolution law, and the maps between scales must satisfy the same simulation condition (Equation~\ref{eq:dyn-diff}).

\begin{figure}[ht]
    \centering
    \includegraphics[width=\linewidth]{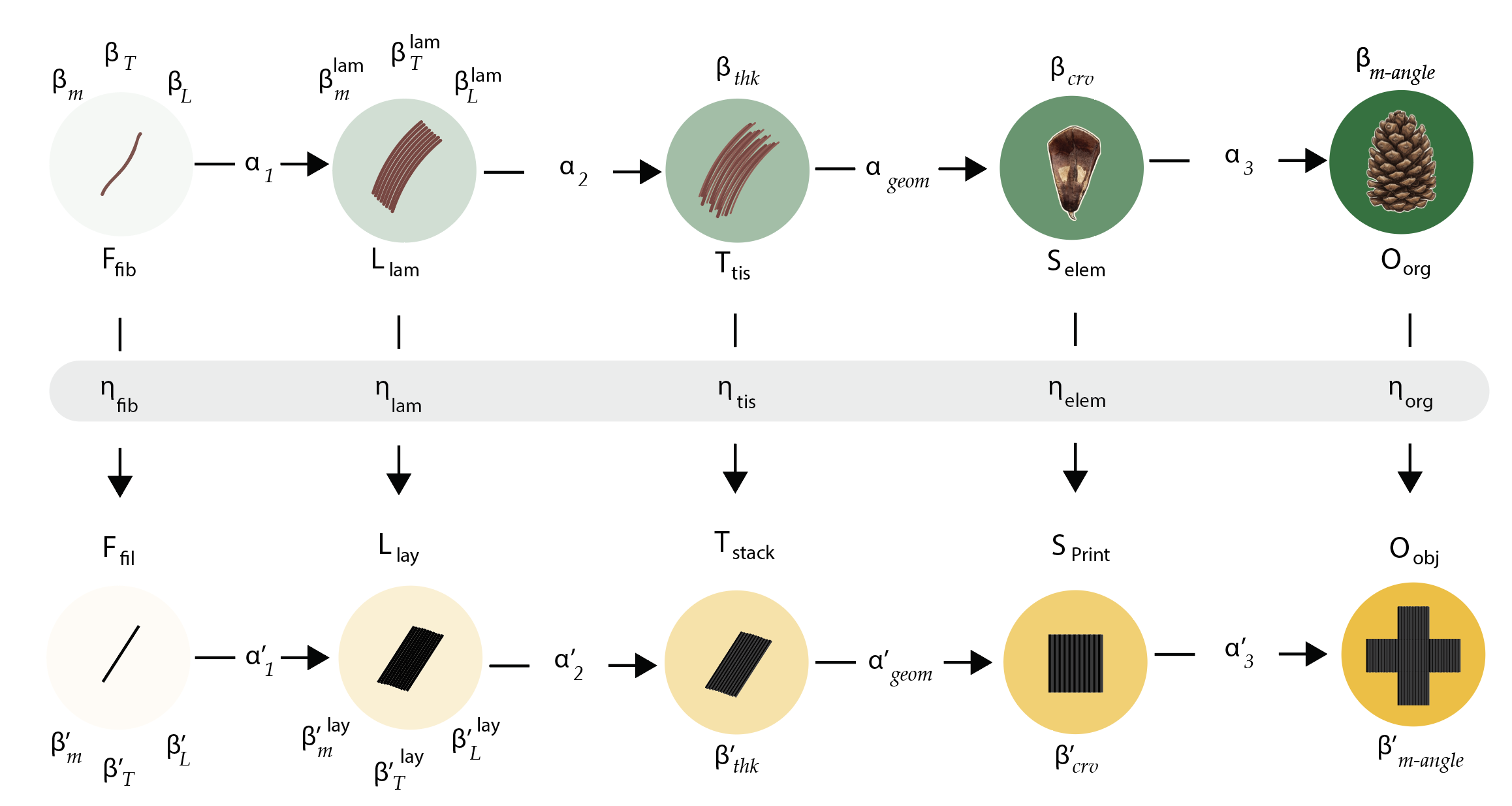}
    \caption{\hlyellow{Biological and engineered hierarchy diagrams and their scale-wise implementation. Top row (green): the biological diagram $D_n:\mathcal{J}\to\Nat$ with assembly morphisms $\alpha$ and reductions $\beta$. Bottom row (yellow): the engineered diagram $D_a:\mathcal{J}\to\Art$, obtained from the implementation functor $\mathcal{F}:\Nat\to\Art$. The vertical arrows are the components $\eta_S$ of a natural transformation $\eta:i_nD_n\Rightarrow i_aD_a$ in $\Dyn$, where $i_n$ and $i_a$ are the inclusions into $\Dyn$. Each square commutes, $\alpha'\circ\eta_S=\eta_T\circ\alpha$, so implementing locally and then assembling gives the same result as assembling and then implementing.}}
    \label{fig:natart}
\end{figure}

The biological hierarchy developed in the preceding section provides a complete stimulus-response description of the pinecone as a running example, from fiber-scale eigenstrain to organ-scale opening angle. 
We now construct the \hlyellow{engineered} hierarchy in $\Art$ introduced in Section~\ref{sec:dyn}. The implementation functor
\begin{equation}
\label{eq:functor}
\mathcal{F}:\Nat\to\Art
\end{equation}
assigns to each biological system an engineered counterpart at the corresponding scale. 
It preserves the state-space organization and the functional form of the evolution law while replacing biological material parameters with engineered ones (Appendix~\ref{appendix}). \hlyellow{The implementation functor relates the two hierarchies scale by scale. As Figure~\ref{fig:natart} indicates, this scale-wise relation is organized as a natural transformation between the biological and engineered hierarchy diagrams, with the functor $\mathcal{F}:\Nat\to\Art$ providing the global translation rule. The formal construction, as an indexed diagram and an implementation natural transformation in $\Dyn$, is given in Appendix~\ref{appendix}.}

We instantiate $\Art$ on 4D printed composites as the engineered analog to the pinecone example formalized in $\Nat$. 
In 4D printing, a flat structure is fabricated by additive manufacturing and subsequently transforms into a target shape under an environmental stimulus~\cite{Tibbits2014}. 
Our target manufacturing platform here is fused filament fabrication (FFF), in which material is deposited layer by layer with controlled raster orientation, producing anisotropic effective properties. 
The structural similarity between the biological and printed hierarchies allows the implementation functor $\mathcal{F}$ to produce a direct, physically realizable translation of nature-derived material design.

On objects, $\mathcal{F}$ assigns to each biological system $S = (X, E, f) \in \Nat$ an engineered system $\mathcal{F}(S) = (X', E', f') \in \Art$ at the corresponding scale:
\begin{equation}
\begin{aligned}
\mathcal{F}(F_{\mathrm{fib}}) &= F_{\mathrm{fil}} & & \text{(filament)} \\
\mathcal{F}(L_{\mathrm{lam}}) &= L_{\mathrm{lay}} & & \text{(printed layer)} \\
\mathcal{F}(T_{\mathrm{tis}}) &= T_{\mathrm{stack}} & & \text{(printed stack)} \\
\mathcal{F}(S_{\mathrm{elem}}) &= S_{\mathrm{print}} & & \text{(printed bilayer)} \\
\mathcal{F}(O_{\mathrm{org}}) &= O_{\mathrm{obj}} & & \text{(printed object)}
\end{aligned}
\end{equation}
On morphisms, $\mathcal{F}$ maps each assembly morphism and reduction in $\Nat$ to a corresponding morphism in $\Art$,
\begin{equation}
\label{eq:functor-morph}
\mathcal{F}(\alpha_i) = \alpha_i', \qquad \mathcal{F}(\beta) = \beta',
\end{equation}
such that all primed morphisms satisfy the simulation condition (Equation~\ref{eq:dyn-diff}) in $\Art$.

The key point is that the engineered hierarchy preserves the compositional mechanics of the biological one, not its material substrate. 
At each scale, the engineered system retains the same dynamical architecture while substituting engineered constitutive and geometric parameters. 
The swelling coefficients $\chi_L$ and $\chi_T$ of the biological fiber are replaced by the actuation coefficients of the printed filament, but the relaxation dynamics retain the same form. 
The assembly morphisms $\alpha'$ perform the same operations as in $\Nat$, averaging, stacking, geometric embedding, and kinematic assembly, but now over printed rather than biological constituents. 
Likewise, the through-thickness mismatch $\Delta\varepsilon_{\mathrm{thk}}$ is generated by contrasts in printed materials and/or raster orientations rather than by differences in microfibril orientation, and the element curvature law takes the same form,
\begin{equation}
\kappa = C_{\mathrm{print}}\,\Delta\varepsilon_{\mathrm{thk}},
\end{equation}
with $C_{\mathrm{print}}$ determined by the printed geometry and effective moduli.

The output of $\mathcal{F}$ is therefore a behavioral target in $\Art$: an engineered hierarchy specified at the level of mechanics and architecture, but not yet at the level of manufacturing instructions. 
That remaining translation is handled by the specification space $\Spec$ in the next section.

\section{Translation from behavioral target to fabrication specification}
\label{sec:spec}

The implementation functor $\mathcal{F}$ delivers a behavioral target $A \in \Art$ that specifies the engineered system design choices, such as materials, raster orientations, and layer thicknesses. 
What remains unresolved is the fabrication specification that physically realizes that target. 
A bilayer with prescribed geometry and material assignment can often be printed through multiple equivalent combinations of print speed, nozzle temperature, infill density, bead width, and layer height. 
The category $\Spec$ formalizes this multiplicity. 
It is process-class specific but machine-agnostic: it captures fabrication intent at the level of the manufacturing method without yet committing to the instruction set of a particular printer. In the present work we instantiate $\Spec$ for additive manufacturing, and specifically for fused filament fabrication (FFF), where behavioral intent is encoded through the sequencing of deposited material and the process parameters assigned to each deposited path.

Each object of $\Spec$ is a fabrication program $\Sigma = (D, \mathcal{P}, \hlyellow{\zeta})$ consisting of a part domain $D \subset \mathbb{R}^3$, an ordered sequence of deposition primitives $\mathcal{P} = \{p_k\}$\hlyellow{, and an annotation map $\hlyellow{\zeta}$ assigning process parameters to each primitive.}. Given the design choices from $\Art$, each primitive carries the remaining process parameters. For FFF, these include print speed, nozzle temperature, bead width, and infill density,
\begin{equation}
  p_k \;\mapsto\; (v_k,\;T_k,\;w_k,\;\rho_k,\;\ldots).
\end{equation}
Morphisms in $\Spec$ are behavior-preserving substitutions of these annotations. 
Composition in $\Spec$ therefore corresponds to chaining process modifications that leave the projected behavioral target unchanged.

Not all process annotations influence behavior equally. Some, such as material identity, layer ordering, and raster orientation, are fixed by the $\Art$-level target because they determine the effective constitutive response. 
Others, such as print speed, nozzle temperature, and, in some cases, infill density or layer height, can vary within admissible ranges without changing the predicted behavior. For any adjustable scalar parameter $\lambda$, we define its process window as the set of values for which the behavioral quantity of interest $Q$ (e.g.\ curvature) remains within a prescribed tolerance $\varepsilon$ of its nominal value:
\begin{equation}
  \mathcal{I}_k := \bigl\{p_k : \lvert Q(p_k) - Q(p_k^{\mathrm{nom}})\rvert \,/\, \lvert Q(p_k^{\mathrm{nom}})\rvert \leq \varepsilon\bigr\}.
\end{equation}
\hlyellow{Each process window $\mathcal{I}_k$ is computed one parameter at a time, holding the others at their nominal values, and identifies the range over which that parameter preserves the behavioral target within tolerance.}

Continuing with our 4D printed hygromorphic example, the $\Art$ target specifies PA6-GF as the active layer (\hlyellow{\mbox{$\phi = 0\degree$}}, thickness $h_1$) and PA612-CF as the passive layer (\hlyellow{\mbox{$\phi = 90\degree$}}, thickness $h_2$). 
The fabrication program $\Sigma_{\mathrm{base}}$ then assigns process parameters to each layer. The active layer requires solid infill ($\rho = 1.0$) to preserve fiber alignment, and the layer resolution $\ell_h = 0.3$\,mm discretizes the target $h_1 = 0.6$\,mm to $n_1 = 2$ passes. Table~\ref{tab:process-baseline} records the process windows and curvature sensitivity for each parameter for this given case.

\begin{table}[ht]
\centering
\caption{Process windows $\mathcal{I}_k$ and curvature sensitivity for the baseline PA6-GF\,/\,PA612-CF bilayer.}
\label{tab:process-baseline}
\renewcommand{\arraystretch}{1.3}
\begin{tabular}{llccc}
\hline
\textbf{Parameter} & \textbf{Layer} & \textbf{Window $\mathcal{I}_k$} & \textbf{Nominal} & $\Delta\kappa/\kappa$ \\
\hline
Nozzle temp $T_k$           & active (PA6-GF)   & $[260,\,290]$\,\degree C            & $275$\,\degree C  & $<1\%$ \\
Nozzle temp $T_k$           & passive (PA612-CF)      & $[280,\,300]$\,\degree C            & $290$\,\degree C  & $<1\%$ \\
Print speed $v_k$           & active             & $[20,\,40]$\,mm/s            & $30$\,mm/s & $<2\%$ \\
Print speed $v_k$           & passive            & $[30,\,60]$\,mm/s            & $45$\,mm/s & $<1\%$ \\
Infill density $\rho_k$     & active             & $1.0$   & $1.0$      & --- \\
Infill density $\rho_k$     & passive            & $[0.8,\,1.0]$                & $1.0$      & ${\sim}8\%$ \\
Layer height $\ell_h$       & both               & $[0.1,\,0.3]$\,mm            & $0.3$\,mm  & $<3\%$ \\
Bed temp $T_{\mathrm{bed}}$ & both               & $90$  & $90$\,\degree C  & --- \\
\hline
\end{tabular}
\end{table}

Given a fabrication specification $\Sigma$, the central question is whether it realizes the intended behavioral target. The projection
\begin{equation}
\label{eq:projection}
\pi:\Spec\to\Art
\end{equation}
answers this by mapping each program to its predicted behavioral target, computing effective moduli ($E_k^{\mathrm{eff}} = \rho_k \cdot E_k^{\mathrm{bulk}}$) and discretizing layer heights to integer pass counts. For the baseline, $\pi(\Sigma_{\mathrm{base}})\cong A_{\mathrm{base}}$, confirming that the chosen process parameters realize the intended behavioral target. The set of all programs that produce the same target $A$ forms the fabrication design space $\pi^{-1}(A)$. Within this space, any substitution of process parameters that stays inside the process windows preserves the behavioral prediction (Appendix~\ref{appendix}).

Once a specification is verified, the last step is to turn it into something a printer can execute. The forgetful functor
\begin{equation}
\label{eq:compilation}
\mathcal{E}:\Spec\to\mathbf{Comp}
\end{equation}
compiles each process parameter into machine syntax (e.g.\ G-code), translating design decisions to print syntax commands. The map $\mathcal{E}$ discards the design reasoning behind each choice and retains only the executable instructions. Different printers correspond to different compilation functors sharing the same $\Spec$ domain.

\section{Computational realization of the framework}
\label{sec:pipeline}

\begin{figure}[h]
    \centering
    \includegraphics[width=1.0\linewidth]{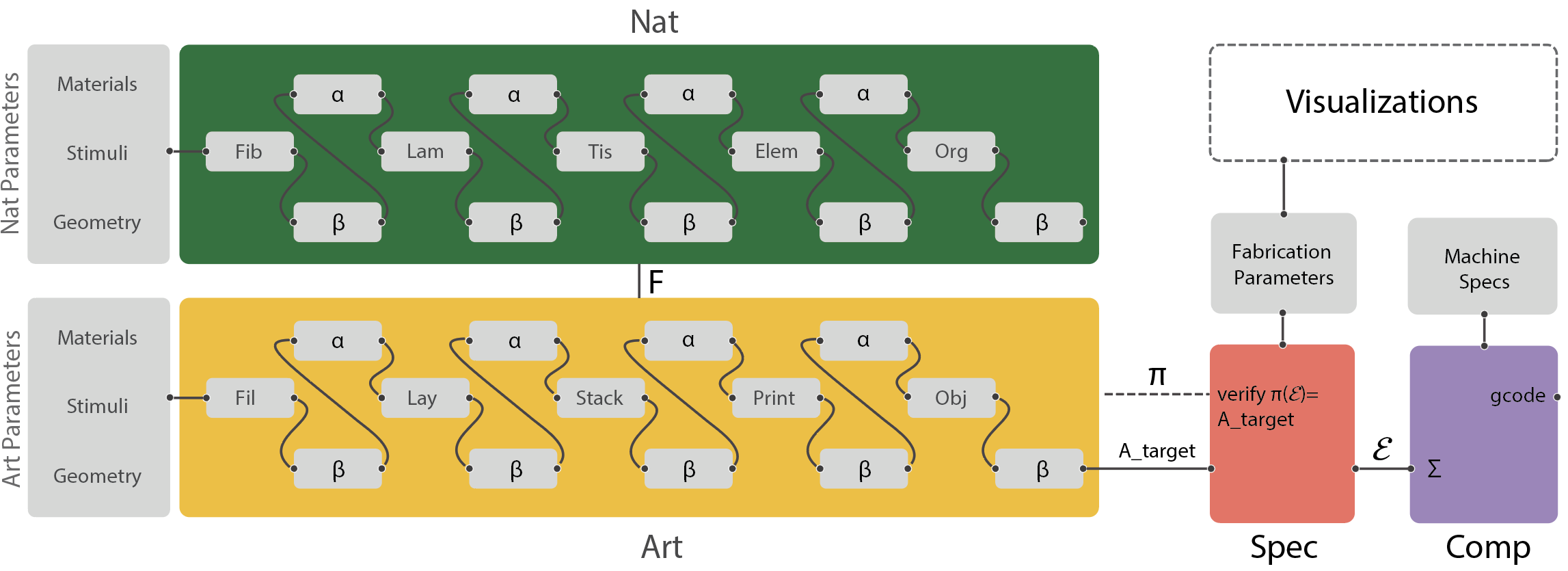}
    \caption{Grasshopper implementation of the categorical pipeline. Material panels encoding biological and engineered material properties feed into separate stimulus collectors. $\Nat$ and $\Art$ evaluate the same governing equations on their respective inputs. $\Art$ outputs the behavioral target $A_{\mathrm{target}}$, which $\Spec$ verifies against fabrication parameters via $\pi$. The verified specification $\Sigma$ is compiled by $\Comp$ into machine-specific G-code. Visualizations are a non-categorical output used for validation.}
    \label{fig:gh}
\end{figure}

The value of the categorical construction is that it is directly executable.
We implement the full pipeline as a modular parametric workflow in Grasshopper (Rhinoceros~3D)~\cite{McNeel2024}. 
Grasshopper's dataflow programming model, in which components are connected by typed wires in a directed graph, provides a natural computational analogue of the categorical diagram. 
Each component realizes a specific categorical map, each wire carries a serialized representation of the corresponding object, and admissible composition is enforced by the graph connectivity itself. 
The result is not merely a visualization of the framework, but an operational pipeline that carries a design from multiscale formalization to fabrication verification and machine execution. Figure~\ref{fig:gh} shows the implemented graph.

The $\Nat$ and $\Art$ branches are evaluated in parallel. 
Each receives the same three classes of inputs, i.e., stimulus data, geometry, and response type, and applies the same hierarchical governing equations. 
They differ only in their input panels. 
$\Nat$ is populated with biological material data describing the natural system (e.g.\ pine cell-wall hygromechanics), whereas $\Art$ is populated with engineered material data describing the printable analog (e.g.\ \hlyellow{CF-PLA} and PLA). \hlyellow{Each component evaluates its governing law in closed form, and material properties enter through panels, one per material, holding approximate parameters taken from datasheets and the literature.}
\hlyellow{In this way, the implementation functor $\mathcal{F}$ is realized computationally as a scale-by-scale evaluation of the same governing laws on engineered material data within a shared compositional architecture.}

\begin{figure}[h]
    \centering    \includegraphics[width=0.8\linewidth]{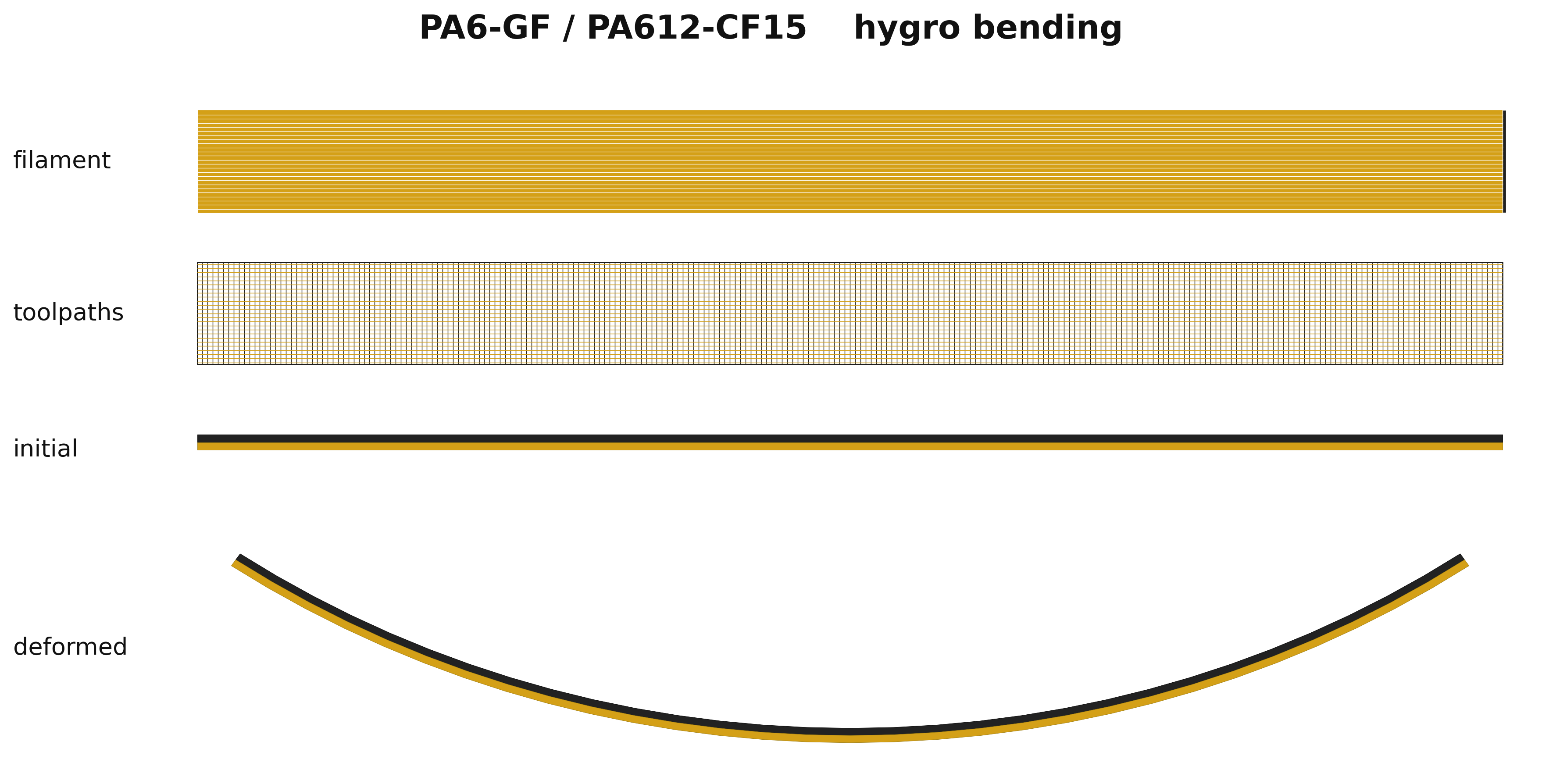}
    \caption{Visualization outputs for the baseline PA6-GF\,/\,PA612-CF hygroscopic bending case. From top to bottom: filament-width rendering of the deposited beads, raster toolpaths with orthogonal orientations for the two layers, initial flat bilayer, and predicted deformed geometry.}
    \label{fig:viz}
\end{figure}

The $\Art$ branch outputs the behavioral target $A_{\mathrm{target}}$, which is passed to $\Spec$ together with the fabrication parameters. 
As these fabrication parameters are adjusted (e.g., infill density, layer height, print speed, or tolerance), the verification condition is updated in real time to indicate whether the current specification remains inside the admissible design space $\pi^{-1}(A_{\mathrm{target}})$. Only verified specifications proceed downstream. 
The compilation layer $\Comp$ then maps the verified specification to machine-specific G-code. A non-categorical visualization component \hlyellow{renders the deformed geometry from the curvature predicted by the pipeline, for pre-fabrication inspection and for visual comparison against the fabricated specimen. Formal verification is handled separately by the projection $\pi$, which checks the specification against the behavioral target}. Figure~\ref{fig:viz} shows an example of a visualization output including toolpaths, fabrication output with bead width, and initial and deformed (actuated) states for a given engineered bilayer design.

Most importantly, the computational graph preserves the modularity of the formalism. 
Local substitutions, such as changing the fiber-scale stimulus law or the tissue-level reduction, propagate automatically through the shared downstream pipeline without requiring the rest of the workflow to be rewritten. 
This is what makes the framework generative in practice: new actuator classes are produced by recombining validated components inside a fixed computational structure.

\section{Case studies: Versatility of the compositional framework}
\label{sec:cases}

The practical value of a compositional framework is that validated components can be recombined to generate new designs without rederivation. 
This section tests that claim directly. 
The pipeline developed above is instantiated on four bilayer actuator designs spanning two stimulus types (hygroscopic and thermal) and two kinematic responses (bending and twisting). Across these four instantiations, only two local components vary: the fiber-scale object $F_i$, which sets the stimulus physics, and the tissue-level reduction $\beta_j$, which selects the kinematic observable. 
All assembly morphisms, the implementation functor $\mathcal{F}$, the projection $\pi$, and the compilation functor $\mathcal{E}$ are shared. This controlled variation isolates the central claim of the paper: new actuator classes arise by recombining validated modules, not by rebuilding the pipeline. \Cref{tab:cases} summarizes the components that vary across the four instantiations.

\paragraph{Baseline: hygroscopic bending.}

The baseline case pairs PA6-GF (active, \hlyellow{\mbox{$\phi = 0\degree$}}) with PA612-CF (passive, \hlyellow{\mbox{$\phi = 90\degree$}}) under a hygroscopic stimulus. The tissue-level reduction $\beta_{\mathrm{thk}}$ extracts the through-thickness strain mismatch, and the Timoshenko bilayer formula yields the bending curvature $\kappa$. This case was developed throughout the preceding sections.

\paragraph{Case I: thermal bending.}

The \hlyellow{multiscale} hierarchy formalized from the pinecone does not depend on the physical origin of the eigenstrain at the fiber scale. Any stimulus that produces anisotropic eigenstrains $(\varepsilon_L^0, \varepsilon_T^0)$ can be substituted into the same architecture. Temperature is such a stimulus~\cite{Timoshenko1925}. The fiber-scale object is replaced by its thermal analogue $F_{\mathrm{fib}}^{\alpha}$, in which the eigenstrain law is $\varepsilon_L^0 = \alpha_L(\theta - \theta_0)$, $\varepsilon_T^0 = \alpha_T(\theta - \theta_0)$. Because the eigenstrain output has identical type, all downstream morphisms remain valid. The reduction $\beta_{\mathrm{thk}}$ is unchanged. The material pair tested is CF-PLA (passive, low CTE, \hlyellow{\mbox{$\phi = 0\degree$}}) and PLA (active, high CTE, \hlyellow{\mbox{$\phi = 90\degree$}}). In the computational pipeline, this case is realized by swapping the stimulus collector from hygroscopic to thermal. No other component is modified. What transfers from the biological system is the hierarchical architecture, not the specific stimulus.

\paragraph{Case II: hygroscopic twisting.}

Hygroscopically driven helical twisting is the dominant opening mechanism in wild wheat awns and other chiral seed dispersal structures, where antisymmetric cellulose fibril orientations convert differential swelling into a twisting moment \cite{Armon2011, Reyssat2009}. The fiber-scale object remains $F_{\mathrm{fib}}$ (hygroscopic). The change is confined to the tissue-level reduction. When a layer's principal swelling axis is oriented at angle $\phi$ to the strip axis, the strain transformation from fiber to strip coordinates introduces an in-plane shear component $\varepsilon_{12} = \tfrac{1}{2}(\varepsilon_L - \varepsilon_T)\sin 2\phi$. In an antisymmetric bilayer with layers at $+\phi$ and $-\phi$, the shear components have equal magnitude but opposite sign, so the through-thickness shear mismatch is
\begin{equation}
  \Delta\varepsilon_{12} = (\varepsilon_L^{(1)} - \varepsilon_T^{(1)})\sin 2\phi.
\end{equation}
This mismatch drives a twist rather than a bend. The reduction $\beta_{\mathrm{twist}}(x_{\mathrm{tis}}) = \Delta\varepsilon_{12}$ satisfies the simulation condition by linearity, and the twist curvature $\tau$ follows from plate theory for antisymmetric laminates. The material pair is PA6-GF (\hlyellow{\mbox{$\phi = +45\degree$}}) and PA612-CF (\hlyellow{\mbox{$\phi = -45\degree$}}). In the computational pipeline, this case is realized by toggling the response from bending to twisting and adjusting the raster orientations. The stimulus collector and all downstream components remain unchanged.

\paragraph{Case III: thermal twisting.}

This case requires no new derivation. The thermal fiber object $F_{\mathrm{fib}}^{\alpha}$, validated in Case~I, and the twist reduction $\beta_{\mathrm{twist}}$, validated in Case~II, compose directly. The resulting shear mismatch is
\begin{equation}
  \Delta\varepsilon_{12}^{\theta} = (\alpha_L^{(1)} - \alpha_T^{(1)})(\theta - \theta_0)\sin 2\phi,
\end{equation}
\hlyellow{This twist could also be obtained by a direct mechanics derivation for this configuration. What is specific to the framework is that the result is produced by composition rather than by a separate derivation.} The composition is guaranteed to yield a valid engineered target: each component morphism satisfies the simulation condition independently, and by closure under composition in $\Dyn$ (Appendix~\ref{appendix}), the composite is again simulation-respecting. The implementation map $\mathcal{F}$ (Appendix~\ref{appendix}) maps the resulting hierarchy to a valid object in $\Art$. In the computational pipeline, this case is realized by selecting the thermal stimulus collector and the twisting response toggle. The pipeline produces a verified specification and executable G-code through the same downstream path as every other case, with no component modified.

\hlyellow{All four cases were modeled and generated as bilayer strips with shared geometry, $L = 100$ mm and $w = 10$ mm, and equal layer thicknesses, $h_1 = h_2 = 0.6$ mm. Table~\mbox{\ref{tab:cases}} summarizes the modeled design and script-level implementation parameters that vary across the four cases.}

\begin{table}[ht]
\centering
\caption{Variations of \hlyellow{modeled} components \hlyellow{and implementation parameters} for the four bilayer actuator designs spanning two \hlyellow{stimulus} types and two kinematic responses.}
\label{tab:cases}
\renewcommand{\arraystretch}{1.35}
\begin{tabular}{lcccc}
\hline
& \textbf{Baseline} & \textbf{Case I} & \textbf{Case II} & \textbf{Case III} \\
\hline
Stimulus          & hygroscopic         & thermal                     & hygroscopic          & thermal \\
Fiber object $F$  & $F_{\mathrm{fib}}$  & $F_{\mathrm{fib}}^{\alpha}$ & $F_{\mathrm{fib}}$  & $F_{\mathrm{fib}}^{\alpha}$ \\
Reduction $\beta$ & $\beta_{\mathrm{thk}}$ & $\beta_{\mathrm{thk}}$ & $\beta_{\mathrm{twist}}$ & $\beta_{\mathrm{twist}}$ \\
Response          & bending  & bending         & twisting    & twisting \\
Active / passive  & PA6-GF / PA612-CF        & PLA / CF-PLA                & PA6-GF / PA612-CF         & PLA / CF-PLA \\
Raster angles     & $0\degree\,/\,90\degree$        & $90\degree\,/\,0\degree$                & $+45\degree\,/\,-45\degree$      & $+45\degree\,/\,-45\degree$ \\
\hline
\end{tabular}
\end{table}

\paragraph{Validation.}

All four specimens were fabricated on a multi-material FFF printer using the G-code generated by $\mathcal{E}(\Sigma)$, with the material pair and raster orientations specified by the verified $\Sigma$ for each case. \hlyellow{Conditioning and testing protocols were kept consistent within each stimulus class.} Thermal actuation specimens (Cases~I and III) were submerged in near boiling water. \hlyellow{H}ygroscopic actuation specimens (Baseline and Case~II) \hlyellow{were dried in an oven and then submerged in room-temperature water} (Supplementary~\ref{supplementary}). In all cases the deformed geometry was empirically measured at equilibrium. Opening angles, mean $\pm$ standard deviation, were \hlyellow{$35.3 \pm 3.1\degree$} (hygroscopic bending, \hlyellow{$n=3$}) and $55.7 \pm 4.2\degree$ (thermal bending, $n=3$), while twisting angles were $41.3 \pm 2.2\degree$ (hygroscopic, $n=4$) and $23.3 \pm 2.5\degree$ (thermal, $n=3$). Notably, no case required manual redesign after G-code generation. The first generated program for each specimen produced the intended actuation mode, with measured responses close to model predictions. \hlyellow{Table~\mbox{\ref{tab:validation}} reports the predicted and measured responses for all four cases, with sample size, standard deviation, and percent error.}

\begin{figure}[!ht]
    \centering
    \includegraphics[width=\linewidth]{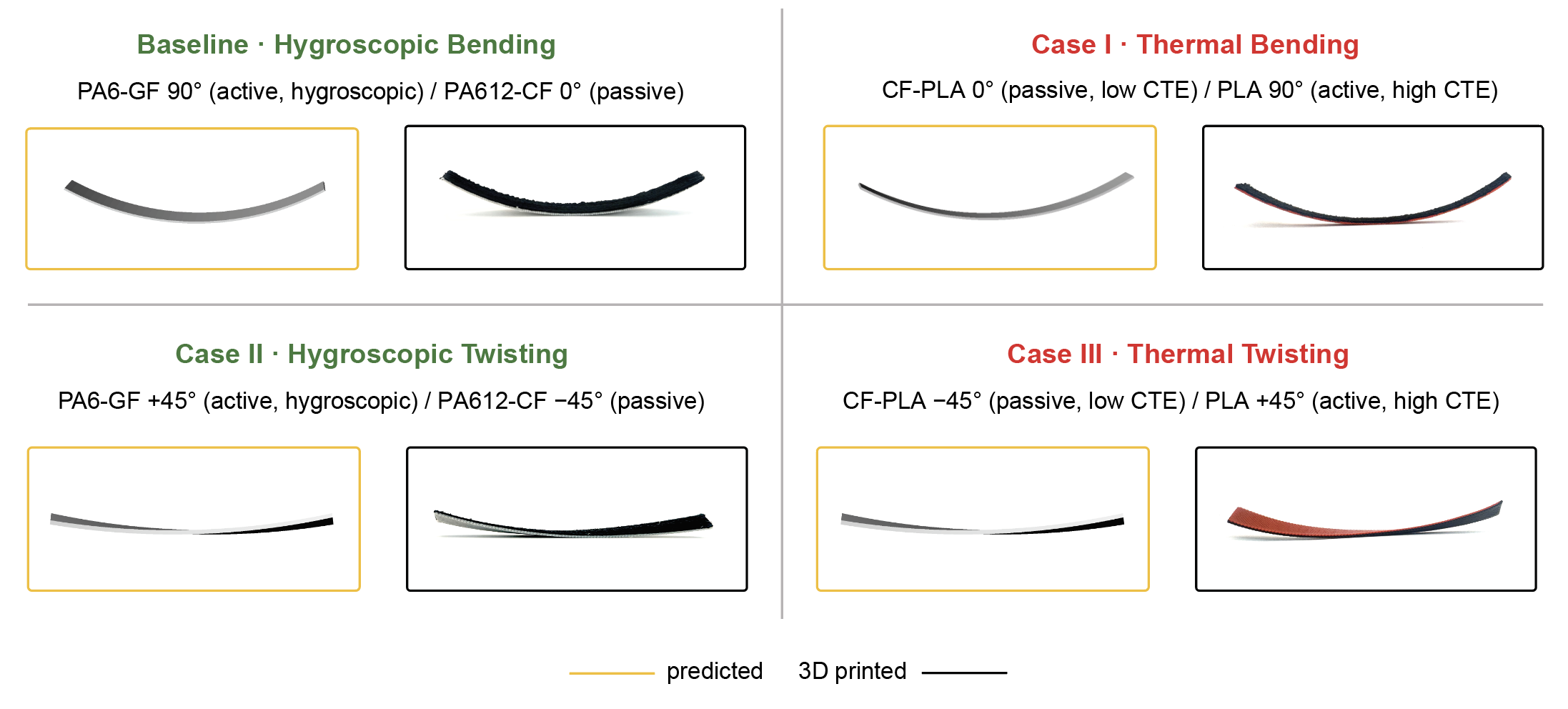}
    \caption{Predicted and observed actuation for all four cases. Each panel shows the deformed geometry computed by the pipeline (left) alongside the experimentally observed response (right). The $2\times 2$ grid is indexed by stimulus type (columns) and kinematic response (rows).}
    \label{fig:cases}
\end{figure}

\begin{table}[ht]
\centering
\caption{\hlyellow{Predicted and measured actuation for the four bilayer cases. Measured values are mean $\pm$ standard deviation across $n$ fabricated specimens. Percent error is relative to the measured mean.}}
\label{tab:validation}
\renewcommand{\arraystretch}{1.35}
\begin{tabular}{lcccc}
\hline
\textbf{Case (stimulus / response)} & $n$ & \textbf{Predicted} & \textbf{Measured (mean $\pm$ s.d.)} & \textbf{\% error} \\
\hline
Baseline (hygroscopic / bending) & 3 & $36.4\degree$ & $35.3 \pm 3.1\degree$ & $3.0\%$ \\
Case I (thermal / bending)       & 3 & $52.0\degree$ & $55.7 \pm 4.2\degree$ & $6.6\%$ \\
Case II (hygroscopic / twisting) & 4 & $41.3\degree$  & $41.3 \pm 2.2\degree$ & $0.1\%$  \\
Case III (thermal / twisting)    & 3 & $23.0\degree$  & $23.3 \pm 2.5\degree$ & $1.4\%$ \\
\hline
\end{tabular}
\end{table}

\hlyellow{All four predictions agree with the measured mean to within one standard deviation.} The agreement confirms that the compositionally generated behavioral targets are physically realized by the fabricated specimens. This closes the loop from categorical formalization to physical actuation. A design that lies within the fabrication design space $\pi^{-1}(A)$ produces the predicted deformation when printed and stimulated.

\section{Conclusion}
\label{sec:conclusion}
This paper introduced an end-to-end compositional framework for nature-derived material design, spanning biological formalization, engineering translation, fabrication specification, and machine execution. 
The category $\Dyn$ provides a common language for stimulus-response systems, while its subcategories $\Nat$ and $\Art$ organize natural and engineered hierarchies within the same formal setting. 
The implementation functor $\mathcal{F}:\Nat\to\Art$ carries a multiscale biological hierarchy into an engineered counterpart. The projection $\pi:\Spec\to\Art$ identifies which fabrication specifications realize a desired behavioral target, and the compilation functor $\mathcal{E}:\Spec\to\mathbf{Comp}$ renders verified specifications executable. 
Instantiated on the pinecone hygromorphic hierarchy and realized through FFF, the framework generated four bilayer actuator classes spanning two stimulus types and two kinematic responses. 
Across all four cases, predicted and measured deformations were in close agreement, showing that the formal pipeline is not only internally consistent but operational from model to specimen.

The central result of the paper is stronger than a proof of descriptive adequacy. 
The framework is generative. 
Its practical value lies in the fact that independently validated components can be recombined to produce new designs without rederiving the full hierarchy. 
The thermal-twisting actuator is the clearest demonstration.
It was not constructed as a separate case from first principles. Instead, it emerged by composing the thermal fiber-scale object validated in Case~I with the twisting tissue-level reduction validated in Case~II, then passing that composite through the same specification, verification, and compilation route as the other designs. In this sense, the fourth actuator class was generated by the compositional structure of the framework itself. 
We emphasize that a single composed case is the minimal proof of this property, not its upper bound. Any library of validated constitutive laws $\{F_i\}$ and \hlyellow{kinematic} reductions $\{\beta_j\}$ closes combinatorially into $|\{F_i\}| \cdot |\{\beta_j\}|$ candidate actuators, each verified by construction.

This distinguishes the present approach from case-by-case biomimetic analogy. 
In conventional bioinspired translation, each new mechanism pairing must be argued and revalidated separately. Here, once a stimulus law and a kinematic reduction are each validated locally, their combinations can be deployed systematically across the same multiscale pipeline. The accessible design space therefore scales with the library of available composable components rather than with the number of manually derived cases. Although the framework was developed to formalize translation from biological material hierarchies, the same logic extends beyond direct biological precedent. Any independently validated components with compatible interfaces can be composed into new design candidates, whether or not a natural prototype exists for that combination.

Read at a higher level of abstraction, this framework belongs to a broader shift in engineering methodology. As systems grow in scale and heterogeneity, correctness at the
system level cannot be recovered by validating each component in isolation. It must be built into the interfaces through which components compose. Formal methods in software, compositional co-design in autonomous systems~\cite{Censi2015,Zardini2021,zardini2023co}, and
model-based verification in cyber-physical engineering all rest on this principle. What is specific to nature-derived material design is that the compositional structure is dictated by the biological hierarchy being translated, and the interfaces carry physical meaning---strain, curvature, opening angle---rather than purely informational content. The categorical framework provides the language in which this physical composition can be made as rigorous as its informational counterparts.

Several extensions follow naturally. The present formulation is deterministic and can be extended to propagate uncertainty in material parameters and fabrication tolerances compositionally through the hierarchy~\cite{Huang2026}. \hlyellow{A fuller account would represent the admissible space jointly, capturing parameter coupling, and would carry process variability through the projection $\pi$ so that admissibility becomes a probabilistic rather than a deterministic condition.} The bilayer actuators studied here are the minimal nontrivial instantiation, but the same formal architecture can accommodate richer geometries, additional scales, and nonlinear constitutive laws by changing local physics while preserving the global pipeline. More broadly, the objects and morphisms of $\mathsf{Dyn}$, $\mathsf{Nat}$, $\mathsf{Art}$, and $\mathsf{Spec}$ constitute a typed, composable physics-driven substrate of exactly the kind that graph- and ontology-driven AI reasoning systems are designed to search over~\cite{Buehler2024,Buehler2025,ghafarollahi2025sciagents,buehler2024accelerating} \hlyellow{, a search the present pipeline exemplifies manually through single-component substitution}. Extending this to multi-agent or neurosymbolic discovery systems can, in principle, enable verified propositions of new fiber-scale stimulus laws, kinematic reductions, and assembly morphisms as candidate objects in $\mathsf{Nat}$. The simulation condition then acts as a type-level filter, realizing verifiable designs as fabrication programs or returning rejected candidates to the proposer with an explicit interface failure. In this reading, compositional structure is not only a language for representation but also a mechanism for verification---the two roles that symbolic and learned systems each supply only partially. 

The broader contribution of this work is therefore methodological. It shows that verified compositional structure can function as a rigorous design principle for translating hierarchical coupled mechanisms from nature into engineered systems, while remaining interoperable with the generative AI systems now being brought to bear on materials
discovery. The commitment that underlies this paper is the same commitment at work in categorical analyses of hierarchical protein materials and music~\cite{Giesa2011,spivak2011ologpnas}, in materiomusic frameworks~\cite{buehler2025selectiveimperfection}, in
graph-based reasoning over scientific knowledge~\cite{Buehler2024,Buehler2025,ghafarollahi2025sciagents,buehler2024accelerating}, and in the biological material-intelligence perspective that motivates this translation~\cite{Marom2025}. Verified compositional translation from nature to fabrication is a concrete
instance of that broader program.

\section*{Author contributions}

\textbf{M.J.B:} Conceptualization, Supervision, Writing - Review \& Editing, Funding acquisition. \textbf{L.M:} Conceptualization, Methodology, Software, Validation, Investigation, Visualization, Writing - Original Draft, Writing - Review \& Editing. \textbf{G.Z:} Methodology, Writing - Review \& Editing. \textbf{S.T:} Methodology.

\section*{Declaration of competing interest}

The authors declare that they have no known competing financial interests or personal relationships that could have appeared to influence the work reported in this paper.

\section*{Supplementary materials}
\label{supplementary}

Supplementary materials include actuation footage of the four case studies (\cref{sec:cases}): \url{baseline.mp4, case1.mp4, case2.mp4, case3.mp4}.

\section*{Data availability}

The data and codes needed to reproduce and evaluate the work of this paper are available in the GitHub repository: \url{https://github.com/lamm-mit/CategoryTheoryDesign}. This includes the Grasshopper scripts, generated G-code, and material parameters. Additional materials are provided as Supplementary Information.

\section*{Acknowledgments}

This work was supported by the MIT Lemelson Engineering Fellowship, Singapore-DSO, and MGAIC.


\appendix
\section{Appendix}
\label{appendix}
\renewcommand{\theequation}{A.\arabic{equation}}
\setcounter{equation}{0}

This appendix collects the formal definitions and proofs underlying the categorical pipeline. 
For readers from mechanics and materials science, we begin with the basic categorical notions used in the paper and then specialize them to the five categories $\Dyn$, $\Nat$, $\Art$, $\Spec$, and $\mathbf{Comp}$, together with the maps $\mathcal{F}$, $\pi$, and $\mathcal{E}$. The main technical fact used throughout is simple: once a local scale transition satisfies the simulation condition, any composition of such transitions also satisfies it.

\subsection{Basic categorical notions}
A \emph{category} $\mathcal{C}$ consists of:
\begin{enumerate}[label=(\roman*)]
    \item a class $\mathrm{Ob}(\mathcal{C})$ of objects;
    \item for every pair of objects $A,B$, a set $\mathrm{Hom}_{\mathcal{C}}(A,B)$ of morphisms $f:A\to B$;
    \item for every object $A$, an identity morphism $\id_A \in \mathrm{Hom}_{\mathcal{C}}(A,A)$;
    \item for every triple $A,B,C$, a composition law
    \begin{equation}
      \mathrm{Hom}_{\mathcal{C}}(B,C)\times \mathrm{Hom}_{\mathcal{C}}(A,B)\to \mathrm{Hom}_{\mathcal{C}}(A,C),
      \qquad (g,f)\mapsto g\circ f,
    \end{equation}
\end{enumerate}
such that:
\begin{equation}
h\circ(g\circ f)=(h\circ g)\circ f
\quad\text{and}\quad
\id_B\circ f=f,\;\; f\circ \id_A=f
\end{equation}
whenever the compositions are defined.

A \emph{subcategory} $\mathcal{D}\subseteq\mathcal{C}$ is obtained by selecting some objects of $\mathcal{C}$ and some morphisms between them, with identities and composition inherited from $\mathcal{C}$. A subcategory is \emph{full} if for every pair of objects $A,B\in \mathrm{Ob}(\mathcal{D})$, one retains all morphisms between them:
\begin{equation}
\mathrm{Hom}_{\mathcal{D}}(A,B)=\mathrm{Hom}_{\mathcal{C}}(A,B).
\end{equation}

A \emph{functor} $F:\mathcal{C}\to\mathcal{D}$ assigns to each object $A\in\mathcal{C}$ an object $F(A)\in\mathcal{D}$ and to each morphism $f:A\to B$ a morphism $F(f):F(A)\to F(B)$ such that
\begin{equation}
F(\id_A)=\id_{F(A)}
\qquad\text{and}\qquad
F(g\circ f)=F(g)\circ F(f).
\end{equation}

Two objects $A$ and $B$ are \emph{isomorphic} if there exist morphisms $f:A\to B$ and $g:B\to A$ such that
\begin{equation}
g\circ f=\id_A,
\qquad
f\circ g=\id_B.
\end{equation}
\subsection{The category \texorpdfstring{$\Dyn$}{Dyn}}
An object of $\Dyn$ is a triple
\begin{equation}
S := (X,E,f),
\end{equation}
where $X$ is a smooth finite-dimensional state manifold, $E$ is a smooth finite-dimensional environment (stimulus) manifold, and
\begin{equation}
f:X\times E\to TX
\end{equation}
is a smooth map with $f(x,e)\in T_xX$ for all $(x,e)$. In coordinates, the dynamics are written
\begin{equation}
\dot{x}=f(x,e).
\end{equation}

A morphism
\begin{equation}
(\alpha,\alpha_E):(X,E,f)\to(Y,F,g)
\end{equation}
in $\Dyn$ is a pair of smooth maps
\begin{equation}
\alpha:X\to Y,
\qquad
\alpha_E:E\to F
\end{equation}
satisfying the \emph{simulation condition}
\begin{equation}
d\alpha_x\big(f(x,e)\big)=g\big(\alpha(x),\alpha_E(e)\big)
\end{equation}
for all $(x,e)\in X\times E$.

For each object $(X,E,f)$, define the identity morphism
\begin{equation}
\id_{(X,E,f)}:=(\id_X,\id_E).
\end{equation}
For composable morphisms
\begin{equation}
(\alpha,\alpha_E):(X,E,f)\to(Y,F,g),
\qquad
(\beta,\beta_F):(Y,F,g)\to(Z,G,h),
\end{equation}
define their composite by
\begin{equation}
(\beta,\beta_F)\circ(\alpha,\alpha_E):=(\beta\circ\alpha,\;\beta_F\circ\alpha_E).
\end{equation}

\textbf{Proposition.} $\Dyn$ is a category.

\textbf{Proof.} We verify identities, associativity, and closure under composition.

For identities,
\[
d(\id_X)_x\big(f(x,e)\big)=f(x,e),
\qquad
\id_E(e)=e,
\]
so $(\id_X,\id_E)$ satisfies the simulation condition.

Associativity follows immediately from associativity of ordinary function composition on the state and environment maps:
\[
(\gamma,\gamma_G)\circ\big((\beta,\beta_F)\circ(\alpha,\alpha_E)\big)
=
\big((\gamma,\gamma_G)\circ(\beta,\beta_F)\big)\circ(\alpha,\alpha_E).
\]

For closure under composition, let
\[
(\alpha,\alpha_E):(X,E,f)\to(Y,F,g),
\qquad
(\beta,\beta_F):(Y,F,g)\to(Z,G,h)
\]
be morphisms in $\Dyn$. Then, by the chain rule and the simulation condition for each map,
\begin{align*}
d(\beta\circ\alpha)_x\big(f(x,e)\big)
&=
d\beta_{\alpha(x)}\!\left(d\alpha_x\big(f(x,e)\big)\right) \\
&=
d\beta_{\alpha(x)}\!\left(g\big(\alpha(x),\alpha_E(e)\big)\right) \\
&=
h\big(\beta(\alpha(x)),\,\beta_F(\alpha_E(e))\big).
\end{align*}
Hence
\[
(\beta\circ\alpha,\;\beta_F\circ\alpha_E)
\]
also satisfies the simulation condition. Therefore $\Dyn$ is a category. \hfill$\square$

\subsection{\texorpdfstring{$\Nat$}{Nat} and \texorpdfstring{$\Art$}{Art} as full subcategories of \texorpdfstring{$\Dyn$}{Dyn}}
$\Nat$ is the full subcategory of $\Dyn$ whose objects are stimulus-response systems arising from natural or biological mechanisms. $\Art$ is the full subcategory whose objects are realized as engineered or fabricated systems. In both cases, the morphisms are precisely the $\Dyn$-morphisms whose source and target both lie in the selected object class.

\textbf{Proposition.} $\Nat$ and $\Art$ are full subcategories of $\Dyn$.

\textbf{Proof.} By construction, every selected object inherits its identity morphism from $\Dyn$, and the composite of two $\Dyn$-morphisms whose source and target lie in the selected class again has source and target in that class. Since all $\Dyn$-morphisms between selected objects are retained, the subcategories are full. \hfill$\square$

\subsection{Reductions and induced observable dynamics}
Not every smooth observable $\beta:X\to Y$ automatically defines a morphism in $\Dyn$. To do so, the retained variables must be \emph{closed} under the dynamics.

Let $S=(X,E,f)\in\Dyn$, and let $\beta:X\to Y$ be a smooth map whose image $\beta(X)$ is an embedded submanifold of $Y$.

We say that $\beta$ is \emph{compatible with the dynamics} if, whenever $\beta(x)=\beta(x')$,
\begin{equation}
d\beta_x\big(f(x,e)\big)=d\beta_{x'}\big(f(x',e)\big)
\quad\text{for all } e\in E.
\end{equation}

\textbf{Proposition.} If $\beta$ is compatible with the dynamics, then there exists a unique smooth induced dynamics
\begin{equation}
f_\beta:\beta(X)\times E\to T\beta(X)
\end{equation}
such that
\begin{equation}
f_\beta(\beta(x),e)=d\beta_x\big(f(x,e)\big)
\quad\text{for all } (x,e)\in X\times E.
\end{equation}
With this induced dynamics, the pair
\begin{equation}
(\beta,\id_E):(X,E,f)\to(\beta(X),E,f_\beta)
\end{equation}
is a morphism in $\Dyn$.

\textbf{Proof.} Define
\[
f_\beta(y,e):=d\beta_x\big(f(x,e)\big)
\quad\text{for any }x\text{ such that }\beta(x)=y.
\]
Compatibility ensures that this is well defined, i.e. independent of the representative $x$. Smoothness follows from smoothness of $\beta$ and $f$ in local coordinates. By construction,
\[
d\beta_x\big(f(x,e)\big)=f_\beta(\beta(x),e),
\]
so $(\beta,\id_E)$ satisfies the simulation condition. Uniqueness is immediate from the defining equation. \hfill$\square$

\medskip

This criterion clarifies the role of reductions in the main text. The \emph{scale-transition reductions} are chosen so that the retained variables are closed under the dynamics and therefore define genuine $\Dyn$-morphisms.
By contrast, several scalar quantities shown in the figures, such as individual strain components, curvature, or mean opening angle, are best understood as \emph{coordinate readouts} or \emph{terminal observables} of already-defined reduced states. They need not, by themselves, carry enough state to define closed dynamics. This distinction does not alter the computational pipeline, but it makes the formal structure precise.

\hlyellow{When the inputs at a scale form a varying family rather than a single object, for example when the fibers within a lamina carry distinct parameters, the input to the averaging morphism $\alpha_1$ is a family indexed over the $N$ fibers. Admissible local schemas may be assigned through a presheaf}
\begin{equation}
\hlyellow{P:\mathcal{T}^{\mathrm{op}}\to\mathbf{Set},}
\end{equation}
\hlyellow{on the specimen tree, where $P(v)$ is the set of admissible material, geometric, or fabrication choices at node $v$, and a global section of $P$ (a family of admissible local choices compatible under the restriction maps) selects one concrete hierarchy. The diagrams used here are such selected sections. The sheaf-theoretic semantics of interconnected dynamical systems~\cite{Schultz2020} is the natural setting for this generalization. The present linear instantiation, with a single object at each scale, does not require it, and we identify the presheaf formulation as the appropriate route to heterogeneous inputs and differing local models in future work.}

\subsection{Static specimen data and assembly morphisms}
Some assembly maps depend on specimen-specific descriptors such as geometry or boundary conditions that remain fixed during actuation. To represent such maps within $\Dyn$, it is convenient to adjoin these descriptors to the state with zero dynamics.

\paragraph{Element-forming map $\alpha_{\mathrm{geom}}$.}

Let $\Omega_{\mathrm{geom}}$ denote the space of fixed geometric and effective-stiffness descriptors, for example
\begin{equation}
\Omega_{\mathrm{geom}}=\{(h_1,h_2,E_1,E_2,L,w)\}.
\end{equation}
Define the augmented tissue object
\begin{equation}
\widetilde{T}_{\mathrm{tis}}
:=
(X_{\mathrm{tis}}\times\Omega_{\mathrm{geom}},\,E_{\mathrm{RH}},\,\widetilde{f}_{\mathrm{tis}})
\end{equation}
with
\begin{equation}
\widetilde{f}_{\mathrm{tis}}((x,\omega),u)=\big(f_{\mathrm{tis}}(x,u),\,0_\omega\big).
\end{equation}
Here $0_\omega$ denotes the zero tangent vector at the fixed descriptor $\omega$.

Let $\Delta\varepsilon_{\mathrm{thk}}(x)$ denote the through-thickness mismatch observable extracted from the tissue state, and let $m_{\mathrm{tis}}(x)$ denote the moisture observable supplied to the element model. Define
\begin{equation}
\alpha_{\mathrm{geom}}(x,\omega)
=
\Big(
C_{\mathrm{geom}}(\omega)\,\Delta\varepsilon_{\mathrm{thk}}(x),\;
\Delta\varepsilon_{\mathrm{thk}}(x),\;
m_{\mathrm{tis}}(x)
\Big).
\end{equation}
Then $\alpha_{\mathrm{geom}}$ maps the augmented tissue state to the element state
\begin{equation}
X_{\mathrm{elem}}=\{(\kappa,\Delta\varepsilon_{\mathrm{thk}},m)\}.
\end{equation}

\textbf{Proposition.} With the element dynamics defined as in the main text, the pair
\begin{equation}
(\alpha_{\mathrm{geom}},\id_{E_{\mathrm{RH}}})
:
\widetilde{T}_{\mathrm{tis}}\to S_{\mathrm{elem}}
\end{equation}
is a morphism in $\Dyn$.

\textbf{Proof.} Along trajectories of $\widetilde{T}_{\mathrm{tis}}$, the descriptor $\omega$ is constant, so
\[
\frac{d}{dt}\Big(C_{\mathrm{geom}}(\omega)\,\Delta\varepsilon_{\mathrm{thk}}(x(t))\Big)
=
C_{\mathrm{geom}}(\omega)\,\frac{d}{dt}\Delta\varepsilon_{\mathrm{thk}}(x(t)).
\]
The remaining retained coordinates evolve according to the induced mismatch and moisture dynamics. This is exactly the element law used in Section~\ref{sec:nat}. Hence the simulation condition is satisfied. \hfill$\square$

\medskip

The important point is that the fixed geometric descriptors are adjoined to the \emph{state}, not the environment. This is what allows the state map $\alpha_{\mathrm{geom}}$ to depend on them while remaining a well-defined morphism in $\Dyn$.

\paragraph{Organ-level map $\alpha_3$.}

Let $A$ denote the space of fixed attachment descriptors (for example clamp location, free length, and orientation), and define the augmented organ object
\begin{equation}
\widetilde{O}_{\mathrm{org}}
:=
(X_{\mathrm{elem}}^{K}\times A^{K},\,E_{\mathrm{RH}},\,\widetilde{f}_{\mathrm{org}})
\end{equation}
with
\begin{equation}
\widetilde{f}_{\mathrm{org}}\big((x_i,a_i)_{i=1}^{K},u\big)
=
\big((f_{\mathrm{elem}}(x_i,u),0_{a_i})\big)_{i=1}^{K}.
\end{equation}

For uniform-curvature elements, let $\kappa_i$ denote the curvature read from $x_i$, and let $L(a_i)$ denote the free length encoded by the attachment descriptor. Define the organ-level configuration map
\begin{equation}
\alpha_3\big((x_i,a_i)_{i=1}^{K}\big)
=
(\theta_i)_{i=1}^{K},
\qquad
\theta_i = L(a_i)\,\kappa_i.
\end{equation}
This gives a configuration object with state space $X_{\mathrm{conf}}\subseteq\mathbb{R}^{K}$.

\textbf{Proposition.} $\alpha_3$ is a $\Dyn$-morphism from $\widetilde{O}_{\mathrm{org}}$ to the induced configuration dynamics on $X_{\mathrm{conf}}$.

\textbf{Proof.} Since each $a_i$ is fixed during actuation, $L(a_i)$ is constant along trajectories. Therefore
\[
\dot{\theta}_i = L(a_i)\,\dot{\kappa}_i,
\]
which is precisely the induced configuration dynamics obtained by differentiating the embedding relation. Hence the simulation condition holds. \hfill$\square$

The mean opening angle
\begin{equation}
\beta_{\mathrm{m\text{-}angle}}((\theta_i)_{i=1}^{K})
=
\frac{1}{K}\sum_{i=1}^{K}\theta_i
\end{equation}
is then a terminal observable on this configuration object.

\subsection{The category \texorpdfstring{$\Spec$}{Spec}}
For the present work, $\Spec$ is defined fiberwise over a fixed part domain and deposition skeleton.

An object of $\Spec$ is a triple
\begin{equation}
\Sigma=(D,\mathcal{P},\hlyellow{\zeta}),
\end{equation}
where:
\begin{itemize}
    \item $D\subset\mathbb{R}^3$ is the part domain,
    \item $\mathcal{P}=(p_1,\dots,p_N)$ is an ordered sequence of deposition primitives,
    \item $\hlyellow{\zeta}$ is an annotation map assigning to each primitive $p_k$ a parameter tuple
    \begin{equation}
      \hlyellow{\zeta}(p_k)=\lambda_k\in\Lambda,
    \end{equation}
    for example
    \begin{equation}
      \lambda_k=(v_k,T_k,w_k,\rho_k,\ell_{h,k},\ldots).
    \end{equation}
\end{itemize}

A morphism
\begin{equation}
\phi:\Sigma\to\Sigma'
\end{equation}
is a primitive-wise annotation update
\begin{equation}
\phi=(\phi_1,\dots,\phi_N),
\qquad
\lambda_k'=\phi_k(\lambda_k),
\end{equation}
such that the underlying part domain and deposition skeleton are unchanged and the projected behavioral target is preserved:
\begin{equation}
D'=D,\qquad \mathcal{P}'=\mathcal{P},\qquad \pi(\Sigma')=\pi(\Sigma).
\end{equation}
Thus, morphisms in $\Spec$ are precisely behavior-preserving substitutions of process annotations.

The identity morphism leaves all annotations unchanged. Composition is defined component-wise:
\begin{equation}
(\psi\circ\phi)_k := \psi_k\circ\phi_k.
\end{equation}

\textbf{Proposition.} $\Spec$ is a category.

\textbf{Proof.} Identity and associativity follow from ordinary function composition on the annotation tuples. For closure, let
\[
\phi:\Sigma\to\Sigma',
\qquad
\psi:\Sigma'\to\Sigma''
\]
be morphisms in $\Spec$. Then
\[
D''=D'=D,
\qquad
\mathcal{P}''=\mathcal{P}'=\mathcal{P},
\]
and, by definition of morphism,
\[
\pi(\Sigma'')=\pi(\Sigma')=\pi(\Sigma).
\]
Hence $\psi\circ\phi$ is again a behavior-preserving substitution, so it is a morphism in $\Spec$. \hfill$\square$

\medskip

For a fixed behavioral target $A\in\Art$, the fiber
\begin{equation}
\pi^{-1}(A)
\end{equation}
is the fabrication design space of all behaviorally equivalent specifications. The process windows defined in Section~\ref{sec:spec} may be viewed as one-parameter slices of this fiber.

\subsection{The category \texorpdfstring{$\mathbf{Comp}$}{Comp}}

An object of $\mathbf{Comp}$ is an executable machine program for a specified printer or machine backend. A morphism
\begin{equation}
r:C\to C'
\end{equation}
is a semantics-preserving program rewrite or translator, meaning that $C$ and $C'$ encode the same deposited geometry and parameter schedule up to machine syntax.

The identity morphism is the trivial rewrite, and composition is ordinary sequential composition of rewrites.

\textbf{Proposition.} $\mathbf{Comp}$ is a category.

\textbf{Proof.} Identity and associativity are inherited from function composition on program rewrites. \hfill$\square$

\subsection{The implementation functor \texorpdfstring{$\mathcal{F}:\Nat\to\Art$}{F: Nat -> Art}}
A rigorous way to formulate $\mathcal{F}$ is as a transport of dynamics along chosen state and environment identifications.

For each object
\begin{equation}
S=(X_S,E_S,f_S)\in\Nat,
\end{equation}
choose an engineered counterpart
\begin{equation}
\mathcal{F}(S)=(X_S',E_S',f_S')\in\Art
\end{equation}
together with diffeomorphisms
\begin{equation}
\iota_S:X_S\to X_S',
\qquad
\jmath_S:E_S\to E_S'
\end{equation}
such that
\begin{equation}
d\iota_{S,x}\big(f_S(x,e)\big)
=
f_S'\big(\iota_S(x),\jmath_S(e)\big)
\quad\text{for all }(x,e)\in X_S\times E_S.
\end{equation}
This condition expresses that the engineered object preserves the state-space organization and evolution law of the biological one up to relabeling and parameter substitution.

For a morphism
\begin{equation}
(\alpha,\alpha_E):S\to T
\end{equation}
in $\Nat$, define
\begin{equation}
\mathcal{F}(\alpha,\alpha_E)
:=
\big(
\iota_T\circ \alpha\circ \iota_S^{-1},
\;
\jmath_T\circ \alpha_E\circ \jmath_S^{-1}
\big).
\end{equation}

\textbf{Proposition.} $\mathcal{F}:\Nat\to\Art$ is a functor.

\textbf{Proof.} Let
\[
(\alpha,\alpha_E):S\to T
\]
be a morphism in $\Nat$, and write
\[
x'=\iota_S(x),\qquad e'=\jmath_S(e).
\]
Then
\begin{align*}
d\!\left(\iota_T\circ\alpha\circ\iota_S^{-1}\right)_{x'}\big(f_S'(x',e')\big)
&=
d\iota_{T,\alpha(x)}
\left(
d\alpha_x\Big(
d(\iota_S^{-1})_{x'}\big(f_S'(x',e')\big)
\Big)
\right) \\
&=
d\iota_{T,\alpha(x)}
\left(
d\alpha_x\big(f_S(x,e)\big)
\right) \\
&=
d\iota_{T,\alpha(x)}
\left(
f_T\big(\alpha(x),\alpha_E(e)\big)
\right) \\
&=
f_T'\big(\iota_T(\alpha(x)),\,\jmath_T(\alpha_E(e))\big).
\end{align*}
Thus $\mathcal{F}(\alpha,\alpha_E)$ satisfies the simulation condition in $\Art$.

Identity preservation is immediate:
\[
\mathcal{F}(\id_S)
=
(\iota_S\circ\id_{X_S}\circ\iota_S^{-1},\;\jmath_S\circ\id_{E_S}\circ\jmath_S^{-1})
=
\id_{\mathcal{F}(S)}.
\]
Composition preservation follows from ordinary composition of conjugated maps:
\[
\mathcal{F}\big((\beta,\beta_F)\circ(\alpha,\alpha_E)\big)
=
\mathcal{F}(\beta,\beta_F)\circ\mathcal{F}(\alpha,\alpha_E).
\]
Hence $\mathcal{F}$ is a functor. \hfill$\square$

\medskip

In the examples studied in this paper, the coordinate types are preserved literally or by simple relabeling, so the maps $\iota_S$ and $\jmath_S$ are typically identities. In that common case, $\mathcal{F}$ reduces to parameter substitution within a fixed hierarchical schema.

As a representative example, the printed element
\begin{equation}
\mathcal{F}(S_{\mathrm{elem}})
=
S_{\mathrm{print}}
=
(X_{\mathrm{elem}},E_{\mathrm{RH}},f_{\mathrm{print}})
\end{equation}
has evolution law
\begin{equation}
  f_{\mathrm{print}}(\kappa,\Delta\varepsilon_{\mathrm{thk}},m;u)
  =
  \begin{pmatrix}
    -C_{\mathrm{print}}\,\tau_{\Delta}^{-1}
      (\Delta\varepsilon_{\mathrm{thk}}-\Delta\varepsilon^{0}_{\mathrm{print}}(m))
    \\[4pt]
    -\tau_{\Delta}^{-1}
      (\Delta\varepsilon_{\mathrm{thk}}-\Delta\varepsilon^{0}_{\mathrm{print}}(m))
    \\[4pt]
    -\tau_m^{-1}(m-m_{\mathrm{eq}}(u))
  \end{pmatrix},
\end{equation}
which has the same symbolic form as $f_{\mathrm{elem}}$, with biological coefficients replaced by engineered ones.

\subsection{\hlyellow{The implementation natural transformation \texorpdfstring{$\eta:i_nD_n\Rightarrow i_aD_a$}{eta: i_n D_n => i_a D_a}}}
\label{sec:nt}

\hlyellow{The functor $\mathcal{F}$ above is the global translation principle. Its action on the pinecone hierarchy is organized by an index category. Let $\mathcal{J}$ be the finite index category}
\begin{equation}
\hlyellow{\mathrm{Fib}\to\mathrm{Lam}\to\mathrm{Tis}\to\mathrm{Elem}\to\mathrm{Org}.}
\end{equation}
\hlyellow{The biological hierarchy is a diagram $D_n:\mathcal{J}\to\Nat$ and the engineered hierarchy a diagram $D_a:\mathcal{J}\to\Art$, with $D_a=\mathcal{F}\circ D_n$. Since a natural transformation requires a common codomain, and $D_n$, $D_a$ land in different subcategories, let}
\begin{equation}
\hlyellow{i_n:\Nat\hookrightarrow\Dyn,\qquad i_a:\Art\hookrightarrow\Dyn}
\end{equation}
\hlyellow{be the inclusion functors. The scale-wise implementation is then a natural transformation}
\begin{equation}
\hlyellow{\eta:i_nD_n\Rightarrow i_aD_a,}
\end{equation}
\hlyellow{whose components are the vertical arrows of Figure~\ref{fig:natart}. For each object $j\in\mathcal{J}$ the component $\eta_j:D_n(j)\to D_a(j)$ (written $\eta_S$ in Figure~\ref{fig:natart}, with $S=D_n(j)$) is a morphism in $\Dyn$, and for every arrow $a:j\to k$ the naturality square}
\begin{equation}
\hlyellow{D_a(a)\circ\eta_j=\eta_k\circ D_n(a)}
\end{equation}
\hlyellow{holds. Each $\eta_j$ is the implementation map carrying the natural object $D_n(j)$ to its engineered counterpart $D_a(j)$. For the pinecone-to-printed-bilayer hierarchy the naturality squares are}
\begin{align}
\hlyellow{\alpha_1'\circ\eta_{\mathrm{Fib}}=\eta_{\mathrm{Lam}}\circ\alpha_1,}&\quad
\hlyellow{\alpha_2'\circ\eta_{\mathrm{Lam}}=\eta_{\mathrm{Tis}}\circ\alpha_2,}\\
\hlyellow{\alpha_{\mathrm{geom}}'\circ\eta_{\mathrm{Tis}}=\eta_{\mathrm{Elem}}\circ\alpha_{\mathrm{geom}},}&\quad
\hlyellow{\alpha_3'\circ\eta_{\mathrm{Elem}}=\eta_{\mathrm{Org}}\circ\alpha_3.}
\end{align}
\hlyellow{Composing them gives}
\begin{equation}
\hlyellow{(\alpha_3'\circ\alpha_{\mathrm{geom}}'\circ\alpha_2'\circ\alpha_1')\circ\eta_{\mathrm{Fib}}
=\eta_{\mathrm{Org}}\circ(\alpha_3\circ\alpha_{\mathrm{geom}}\circ\alpha_2\circ\alpha_1),}
\end{equation}
\hlyellow{so implementation commutes with hierarchical assembly. Each naturality square is the simulation-condition identity established above in defining $\mathcal{F}$ on morphisms.}

\subsection{The projection functor \texorpdfstring{$\pi:\Spec\to\Art$}{pi: Spec -> Art}}

On objects, $\pi$ maps each fabrication specification
\begin{equation}
\Sigma=(D,\mathcal{P},\hlyellow{\zeta})
\end{equation}
to the engineered behavioral target
\begin{equation}
\pi(\Sigma)=A_\Sigma\in\Art
\end{equation}
obtained by computing the effective mechanical consequences of the process annotations. In the FFF setting of this paper, this includes quantities such as effective moduli, raster-induced anisotropy, and discretized layer counts.

On morphisms, if
\begin{equation}
\phi:\Sigma\to\Sigma'
\end{equation}
is a morphism in $\Spec$, then by definition
\begin{equation}
\pi(\Sigma')=\pi(\Sigma).
\end{equation}
Hence $\pi$ maps $\phi$ to the identity morphism on the common target:
\begin{equation}
\pi(\phi):=\id_{\pi(\Sigma)}.
\end{equation}

\textbf{Proposition.} $\pi:\Spec\to\Art$ is a functor.

\textbf{Proof.} Identity preservation is immediate:
\[
\pi(\id_\Sigma)=\id_{\pi(\Sigma)}.
\]
For composable morphisms
\[
\phi:\Sigma\to\Sigma',
\qquad
\psi:\Sigma'\to\Sigma'',
\]
we have
\[
\pi(\phi)=\id_{\pi(\Sigma)},
\qquad
\pi(\psi)=\id_{\pi(\Sigma')},
\]
and since $\pi(\Sigma)=\pi(\Sigma')=\pi(\Sigma'')$,
\[
\pi(\psi\circ\phi)
=
\id_{\pi(\Sigma)}
=
\id_{\pi(\Sigma'')}\circ \id_{\pi(\Sigma)}
=
\pi(\psi)\circ\pi(\phi).
\]
Thus $\pi$ preserves identities and composition. \hfill$\square$

\medskip

In this sense, $\pi$ forgets fabrication syntax while retaining only the predicted behavioral content.

\subsection{The compilation functor \texorpdfstring{$\mathcal{E}:\Spec\to\mathbf{Comp}$}{E: Spec -> Comp}}
For a fixed machine backend, the compilation functor
\begin{equation}
\mathcal{E}:\Spec\to\mathbf{Comp}
\end{equation}
maps each specification
\begin{equation}
\Sigma=(D,\mathcal{P},\hlyellow{\zeta})
\end{equation}
to an executable program
\begin{equation}
\mathcal{E}(\Sigma)=C_\Sigma
\end{equation}
by translating each primitive and annotation tuple into the corresponding machine instructions.

If
\begin{equation}
\phi:\Sigma\to\Sigma'
\end{equation}
is a morphism in $\Spec$, then $\mathcal{E}(\phi)$ is the corresponding program rewrite that updates the instruction blocks associated with the changed annotations while preserving the primitive ordering.

\textbf{Proposition.} $\mathcal{E}$ is a functor.

\textbf{Proof.} Identity preservation holds because leaving every annotation unchanged leaves every generated instruction block unchanged:
\[
\mathcal{E}(\id_\Sigma)=\id_{\mathcal{E}(\Sigma)}.
\]
Composition preservation follows because compiling a successive pair of annotation substitutions is the same as compiling their composite:
\[
\mathcal{E}(\psi\circ\phi)=\mathcal{E}(\psi)\circ\mathcal{E}(\phi).
\]
Hence $\mathcal{E}$ is a functor. \hfill$\square$

\medskip

Different printers or machine languages correspond to different compilation functors defined on the same machine-agnostic source category $\Spec$.

\bibliographystyle{naturemag}
\bibliography{export}

\end{document}